\documentclass[useAMS,usenatbib]{mn2e}
\usepackage{amsmath,astrojournals,fleqn,graphicx,txfonts}

\newcommand {\eq}{{\,=\,}}
\newcommand {\ms}{m_{\mathrm{s}}}
\newcommand {\rs}{r_{\mathrm{s}}}
\newcommand {\vs}{{\upsilon_{\mathrm{s}}}}
\newcommand {\rperi}{r_{\mathrm{peri}}}
\newcommand {\tinfall}{t_{\mathrm{infall}}}
\newcommand {\msun}{\mathrm{M}_\odot}
\newcommand {\B}[1] {\boldsymbol{#1}}

\title[Weakening dark-matter cusps by baryonic infall]
      {Weakening dark-matter cusps by clumpy baryonic infall}

\author[David R.~Cole, Walter~Dehnen, Mark I.~Wilkinson]
       {David R.~Cole, Walter Dehnen, Mark I.~Wilkinson%
\thanks{Email: david.cole@astro.le.ac.uk, walter.dehnen@astro.le.ac.uk,
  mark.wilkinson@astro.le.ac.uk}\\
  Theoretical Astrophysics Group,
  Department of Physics \& Astronomy,
  University of Leicester,
  Leicester, LE1~7RH
}

\date{Accepted .
      Received ;
      }
 
\pagerange{\pageref{firstpage}--\pageref{lastpage}}
\pubyear{2011}
 
\begin{document}

\maketitle
\label{firstpage}

\begin{abstract}
  We consider the infall of a massive clump into a dark-matter halo as a
  simple and extreme model for the effect of baryonic physics (neglected in
  gravity-only simulations of large-scale structure formation) on the
  dark-matter. We find that such an infalling clump is extremely efficient in
  altering the structure of the halo and reducing its central density: a clump
  of 1\% the mass of the halo can remove about twice its own mass from the
  inner halo and transform a cusp into a core or weaker cusp. If the clump is
  subsequently removed, mimicking a galactic wind, the central halo density is
  further reduced and the mass removed from the inner halo doubled. Lighter
  clumps are even more efficient: the ratio of removed mass to clump mass
  increases slightly towards smaller clump masses. This process is the more
  efficient the more radially anisotropic the initial dark-matter
  velocities. While such a clumpy infall may be somewhat unrealistic, it
  demonstrates that the baryons need to transfer only a small fraction of
  their initial energy to the dark matter via dynamical friction to explain
  the discrepancy between predicted dark-matter density profiles and those
  inferred from observations of dark-matter dominated galaxies.
\end{abstract}

\section{Introduction}
\label{sec:intro}
The success of the $\Lambda$CDM paradigm of cosmological structure formation
in reproducing the observed structure of the universe on scales
$\gtrsim1\,$Mpc is now well-established \citep*[see e.g.][and references
  therein]{SpringelFrenkWhite2006}. However, on the scale of individual
galaxies ($\lesssim 100$kpc) the density profiles of dark matter haloes pose a
potentially significant problem. The form of these profiles in the absence of
baryonic physics has been extensively studied by means of numerical
simulations. \cite{DubinskiCarlberg1991} and \cite*{NavarroFrenkWhite1997}
showed that haloes on a range of mass scales have similar profiles, with the
density in the innermost regions exhibiting an $\rho\propto r^{-1}$ cusp. More
recent work has found that the haloes can be better represented by profiles
with either slightly shallower inner cusp slopes
\citep{DehnenMcLaughlin2005,NavarroEtAl2010} or continuously varying slope,
for example the \cite{EinastoEinasto1972} profile \citep{NavarroEtAl2004}.

However, there is mounting observational evidence that galaxies occupy
dark-matter haloes with almost uniform density cores with the strongest
evidence coming from low surface brightness galaxies \citep[see][for a recent
  review]{deBlok2009}. Although there have been fewer such studies in high
surface brightness spirals, recent work by \cite{SpanoEtAl2008} has shown that
in these galaxies also cored haloes are favoured. Even in low-luminosity,
pressure supported systems such as the dwarf spheroidal satellite galaxies
surrounding the Milky Way, there is circumstantial evidence that their haloes
are not cusped \citep{GilmoreEtAl2007}.

It is important to remember however, that the ignorance of baryonic physics in
the aforementioned simulations constitutes a significant limitation with
regard to any discussion of the inner halo profiles. The inclusion of baryon
physics is widely recognised as a crucial, albeit extremely technically
challenging prerequisite for further progress in understanding galaxy
formation.

There are multiple ways in which the baryons can affect the density profile of
a dark matter halo. First, a cloud of gas which initially extends throughout a
cusped dark matter halo can dissipate energy by radiation and contract to the
centre of the halo. As the dark matter responds to the deeper gravitational
potential, this in turn leads to a steepening of the dark matter cusp, known
as adiabatic contraction \citep{BlumenthalEtAl1986}. This is not necessarily
the end of the story, however, as baryons also have the ability to generate
mass outflows driven by stellar winds and supernovae produced as a result of
star formation. Depending on the efficiency of star formation, such processes
can expel a significant fraction of the baryonic mass from the central regions
of a galaxy, resulting in a large-scale rearrangement of the dark matter.
Using semi-analytic arguments, \cite{GnedinZhao2002} found that when stellar
mass loss was preceded by adiabatic contraction, the resulting halo
distribution was almost unchanged from its original cusped profile.
\cite{ReadGilmore2005}, however, subsequently showed that repeated episodes of
adiabatic contraction followed by rapid mass expulsion could give rise to a
reduction in the central density and hence produce cored haloes. This occurs
because although the initial infall/outflow produces the same mass density
profile, the velocity structure of the dark matter halo after the outflow is
biased towards radial orbits. As a result, subsequent events are able to lower
the inner density more easily.

Another way in which the baryonic component of a galaxy can transfer energy to
the dark matter halo is by means of a stellar bar. \cite{WeinbergKatz2002}
proposed that the rotation of bars is decelerated by the exchange of energy
and angular momentum with dark-matter particles on orbits in resonance with
the bar's rotation. While further work by \cite{Athanassoula2002,
  Athanassoula2003} confirmed this conclusion, the size required for a stellar
bar to significantly change the mass distribution in the inner regions of a
dark-matter halo was found to be much larger than those observed in barred
galaxies. Hence, this mechanism is thought to be of limited importance for the
majority of dark-matter haloes, though the deceleration process is certainly
affecting the evolution of galactic bars.

However, galaxy formation also involves violent processes, where baryonic
inflow is neither smooth nor adiabatic. Gas accretion is likely to occur
during galaxy mergers, when it takes the form of clumpy infall rather than the
slow contraction of a smooth cloud. Although a baryon clump falling into the
centre of a dark halo will add to the gravitational potential there (and thus
increase the binding energy of the dark halo), it can, during this process,
lose its orbital energy via dynamical friction.
\cite*{ElZantShlosmanHoffman2001} showed that the energy thus gained
(i.e.\ binding energy lost) by the dark matter can produce an observable
impact on the dark matter density profile. \cite{ElZantEtAl2004} and
\cite{NipotiEtAl2004} extended this work by performing $N$-body simulations of
galaxy clusters, where the infalling galaxies play the role of clumps.  They
found that the initial dark-matter cusp can be softened through the transfer
of energy from the baryonic clumps to the dark-matter though the overall
density profile, including the baryonic component, remained cusped.

Attempts have also been made to add full baryonic physics to the studies of
sinking clumps mainly based on the results of cosmological simulations, which
try to model the effect of cooling, metal enrichment and supernova feedback
(e.g.\ \citealt{GnedinEtAl2004}; \citealt{RomanoDiazEtAl2008};
\citealt*{PedrosaTisseraScannapieco2009};
\citealt*{JohanssonNaabOstriker2009}). These simulations confirmed the
transfer of energy and angular momentum from baryons to the dark-matter, while
the results on the dark-matter density reduction were conflicting. This is
presumably because of varying amounts, depending on the details of the
respective model, of contraction owed to the additional gravitational pull
from the accreted baryons.

Despite these promising attempts, a truly realistic modelling of baryonic
physics is still beyond contemporary simulation techniques, not least because
important baryonic processes, such as re-ionisation as well as primordial and
ordinary star-formation, are themselves not sufficiently
understood. Therefore, it is important to understand more quantitatively the
purely stellar dynamical aspect of this mechanism, which alone affects the
dark-matter distribution. Some progress towards this goal has been made
recently by \cite{JardelSellwood2009} and \cite{GoerdtEtAl2010}.  However, a
complete understanding of the pure stellar dynamical problem is still missing,
but seems essential before attempting to interpret the results of simulations
which include baryons. In the present paper, we build on this previous work to
explore the impact of clumpy baryonic infall more broadly. We consider more
realistic initial conditions. In particular, we focus on clumps initially on
parabolic orbits, which may become bound to a halo during a merger event, and
haloes with anisotropic velocity distributions, the expected situation to
obtain within the hierarchical structure formation scenario of CDM.

The outline of the paper is as follows. In Section~\ref{sec:analytic} we
consider analytical estimates for the damage done to the halo by the energy
transfer from the satellite orbit. Section~\ref{sec:model} gives our modelling
approach for the $N$-body simulations, while Sections~\ref{sec:decay} and
\ref{sec:halo:change} describe the resulting orbital decay and typical changes
induced in the simulated haloes.  Sections~\ref{sec:vary:beta} and
\ref{sec:beta:r} discuss the effect of halo velocity anisotropy on both
orbital decay and damage to the halo, while Section~\ref{sec:vary:ms:rs}
considers the effect of satellite mass and size. In Section~\ref{sec:removal}
we demonstrate the effect of removing the accreted clump, corresponding to a
galactic wind subsequent. Finally, Sections~\ref{sec:summary} and
\ref{sec:discuss} summarise and discuss our findings, respectively.

\section{Theoretical arguments}
\label{sec:analytic}
\citeauthor{Chandrasekhar1943}'s (1943) dynamical friction formula for systems
with a Maxwellian velocity distribution of dispersion $\sigma$ \citep[eq.~8.7
  of][]{BinneyTremaine2008}
\begin{equation} \label{eq:dynfr}
  \frac{\mathrm{d}\B{\upsilon}_{\mathrm{s}}}{\mathrm{d}t}
  \simeq -  \hat{\B{\upsilon}}_{\mathrm{s}}\,
  \frac{4\pi G^2\ms\,\rho\ln\Lambda}{\upsilon_{\mathrm{s}}^2}
  \left[\mathrm{erf}(x)-\frac{2x}{\sqrt{\pi}}\mathrm{e}^{-x^2}
    \right]_{x=\vs/\sqrt{2}\sigma}
\end{equation}
($\ln\Lambda$ is the Coulomb logarithm, $\ms$ and $\B{\upsilon}_{\mathrm{s}}$
are the mass and velocity of the clump or satellite, and $\rho$ is the mass
density of dark-matter particles) shows that the deceleration is proportional
to $m_s$, such that the time for the orbit to decay
$t_{\mathrm{infall}}\propto \ms^{-1}$. In particular, for this orbital decay
to occur within (less than) a Hubble time, a mass of $m_s\sim10^{6-8}M_\odot$,
depending on the size of the dark-matter halo, is required. Chandrasekhar's
formula also suggests that the drag force is strongest for small $\vs$
(because this increases the interaction time between perturber and dark-matter
particles) and/or for large $\rho$.

However, the formula cannot be used to assess the effect the infalling clump
has on the dark matter. A simple estimate for the mass removed from the inner
parts of the dark-matter halo can be obtained from the following argument
originally due to \cite[][preprint version]{GoerdtEtAl2010}. Assuming a
circular orbit for the perturber, the specific energy lost when sinking from
radius $r+\delta r$ to $r$ is
\begin{equation}
  \label{eq:deltaA}
  \delta \varepsilon_{\mathrm{s}} = \frac{\mathrm{d}}{\mathrm{d}r}
  \left[\frac{GM(r)}{2r}+\Phi(r)\right] \delta r =
  2\pi\,G\,r\left(\frac{\bar{\rho}}{3} + \rho\right) \delta r
\end{equation}
with $\bar{\rho}(r)$ the mean density interior to radius $r$. Assuming this
energy is injected into the spherical shell at radius $r$, each dark-matter
particle at that radius gains specific energy $(\delta\varepsilon_{\mathrm{s}}
/\delta r) (\ms/4\pi\rho r^2)$. A density core forms and the sinking of the
clump stalls \citep{ReadEtAl2006} as soon as this energy equals the specific
kinetic energy of each particle, which may be estimated as
$\upsilon_{\mathrm{circ}}^2/2=GM(r)/2r$. With (\ref{eq:deltaA}), this yields
(with $\gamma=-\mathrm{d}\ln\rho/\mathrm{d}\ln r$)
\begin{equation}
  \label{eq:Mrm}
  M(r) \sim \left[1+\frac{\bar{\rho}(r)}{3\rho(r)}\right]
  \,\ms \simeq \frac{4-\gamma}{3-\gamma}\,\ms.
\end{equation}
This argument suggests (i) that the mass ejected by the perturber is
comparable to its own mass, and (ii) that the density core which forms in
response to the heating induced by the sinking baryonic clump has radius
comparable to that at which the originally enclosed mass equals $\ms$.

Strictly speaking, this argument only applies to circular orbits, which are
not very realistic, and the assessment of the heating required to turn a cusp
into a core is rather crude. We now present a more quantitative estimate based
on the exact energy difference between initial and final halo and on the
assumption that the orbital energy lost by the clump is absorbed by (the inner
parts of) the halo.

\begin{figure*}
  \centerline{
    \resizebox{84mm}{!}{\includegraphics{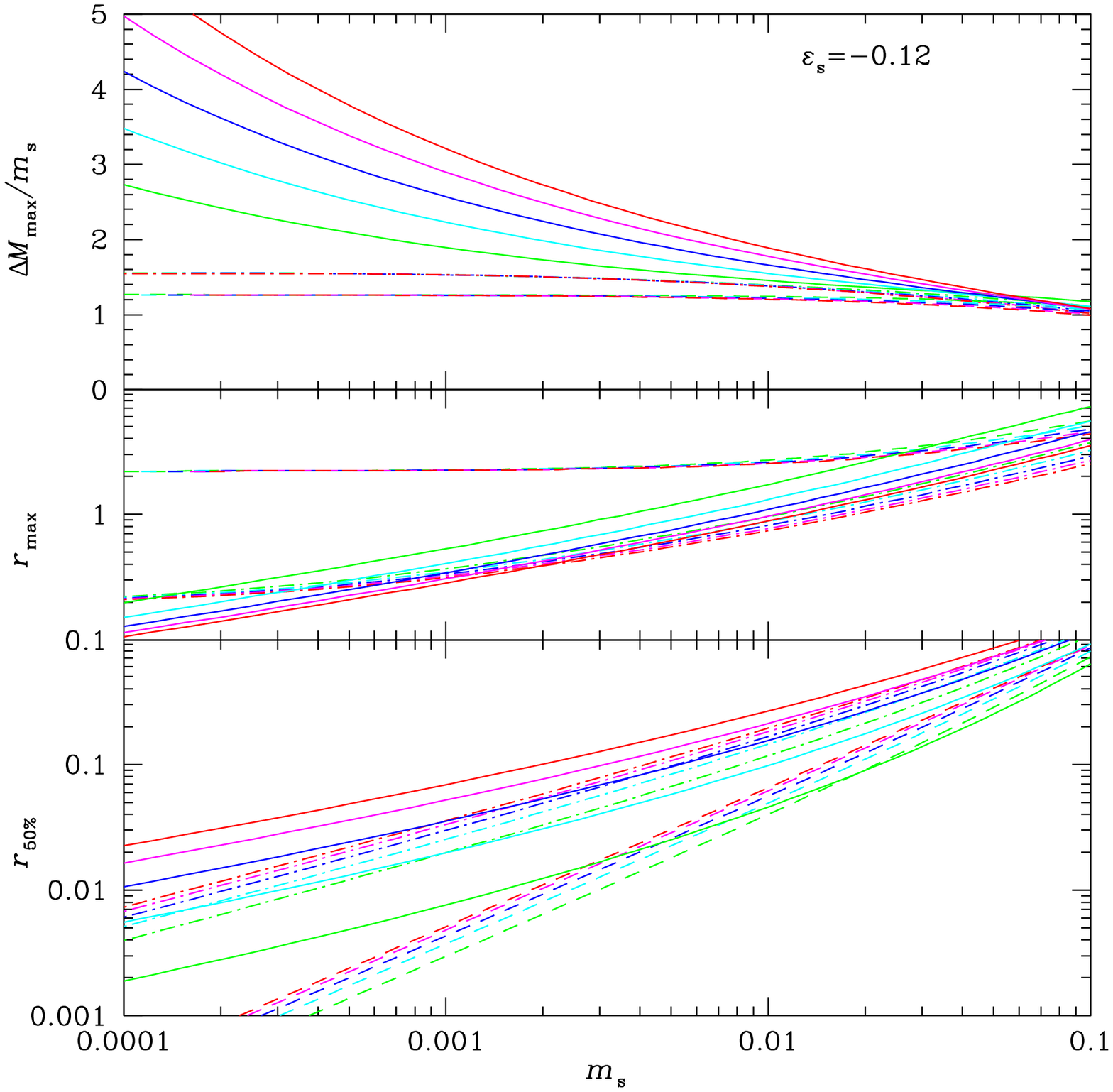}}
    \hfil
    \resizebox{84mm}{!}{\includegraphics{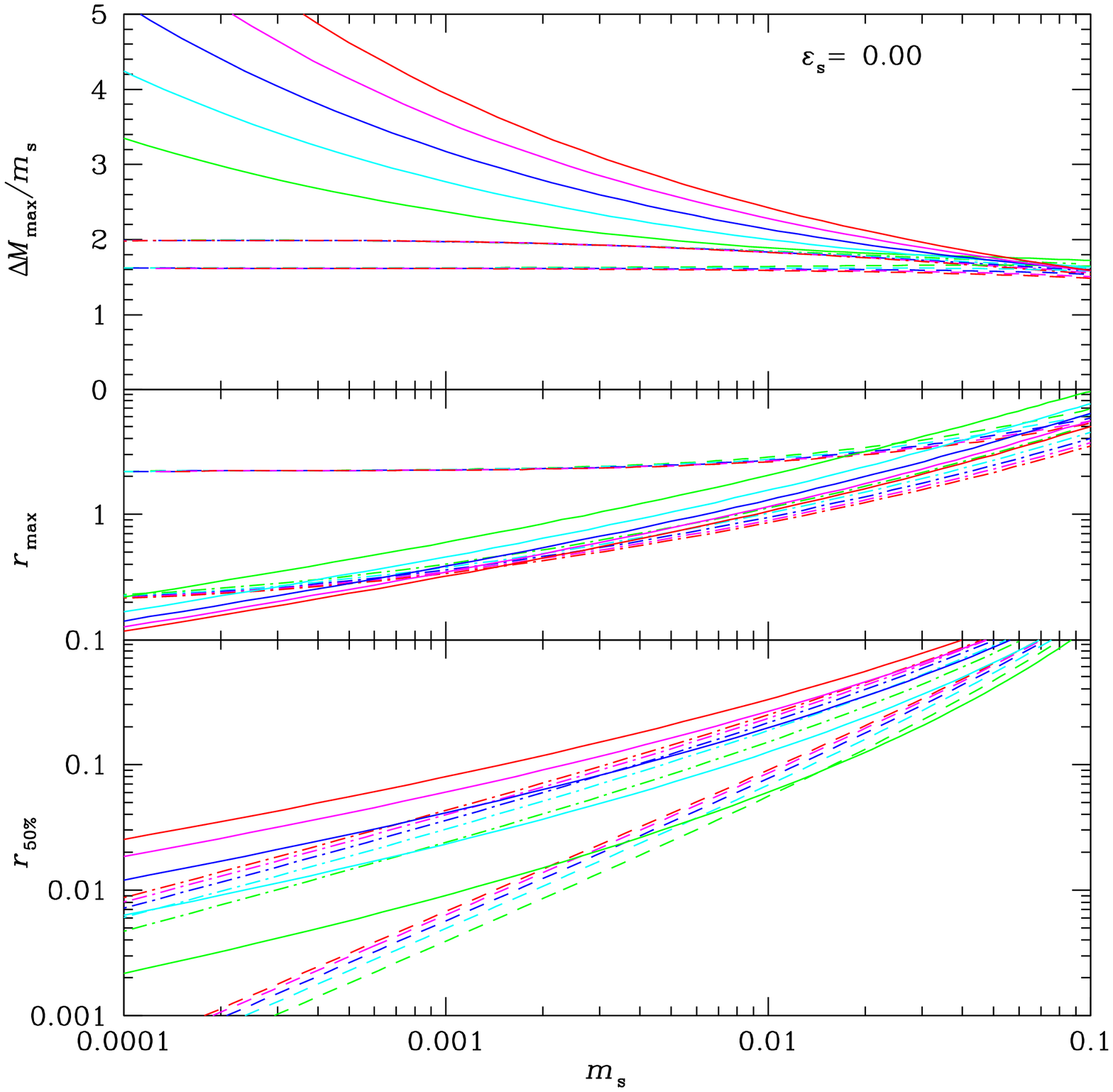}}
  }
  \caption{\label{fig:dM} Predictions, using
    equation~(\ref{eq:energy:balance}), for the ratio of the maximum excavated
    mass to the mass of the infalling clump (\emph{top}), the radius at which
    this occurs (\emph{middle}), and the radius inside of which half the mass
    has been removed (\emph{bottom}; in units of halo scale radius) as
    function of the mass of the accreted clump. The clump was assumed to be
    initially either on a bound orbit with energy equal to that of the
    circular orbit at halo half-mass radius (\emph{left}) or on a parabolic
    orbit (\emph{right}). The initial halo is modelled to have density
    distribution~(\ref{eq:rho}) also used in our simulations, while for the
    final halo we used the models of equation~(\ref{eq:B}) (\emph{solid}) or
    (\ref{eq:A}) (\emph{dashed}: $\eta=0.9$ or \emph{dash-dotted}:
    $\eta=1.5$). The different lines correspond to inner density slopes for
    the final halo between $\gamma_0=0$ (\emph{red}) and $\gamma_0=0.4$
    (\emph{green}), while along each line the scale radii $r_0$ of the final
    model are varied (and $\ms$ obtained from equation~\ref{eq:energy:balance}).
  }
\end{figure*}
Assuming spherical symmetry, let $M_{\mathrm{i}}(r)$ and $M_{\mathrm{f}}(r)$
denote the cumulative mass profiles of the initial and final halo. At large
radii the halo is hardly altered, i.e.\ $M_{\mathrm{f}}(r)\approx
M_{\mathrm{i}}(r)$, while at small radii the halo has been heated resulting in
an expansion and hence $M_{\mathrm{f}}(r)<M_{\mathrm{i}}(r)$. A quantitative
relation between the change in $M(r)$ and the mass $\ms$ of the clump can be
obtained by considering the total energy budget. By virtue of the virial
theorem, the total energy of the initial equilibrium halo is half its
potential energy $V_{\mathrm{i}}$, to which the kinetic energy of the clump
and the interaction energy between clump and halo must be added to obtain the
total energy of the initial state (neglecting the clump self-energy)
\begin{equation}
  E_{\mathrm{i}}=\frac{\ms}{2}\left(\upsilon_{\mathrm{i}}^2
  +\Phi_{\mathrm{i}}(r_{\mathrm{i}})\right) + \frac{1}{2}V_{\mathrm{i}}.
\end{equation}
with $r_{\mathrm{i}}$ and $\upsilon_{\mathrm{i}}$ the initial radius and speed
of the clump and $\Phi_{\mathrm{i}}(r)$ the potential due to the initial
halo. Expressing this in terms of the initial specific orbital energy
$\varepsilon_{\mathrm{s}}\equiv \tfrac{1}{2}\upsilon_{\mathrm{i}}^2 +
\Phi_{\mathrm{i}}(r_{\mathrm{i}})$ of the clump and the cumulative mass
profile gives
\begin{equation}
  E_{\mathrm{i}} = \ms\left[\varepsilon_{\mathrm{s}}-\frac{1}{2}
  \Phi_{\mathrm{i}}(r_{\mathrm{i}}) \right] - \frac{G}{4} \int_0^\infty
  \frac{M^2_{\mathrm{i}}(r)}{r^2}\,\mathrm{d}r,
\end{equation}
where we have used the relation $V{\,=\,}-\tfrac{G}{2}\int_0^\infty
M^2(r)\,r^{-2}\,\mathrm{d}r$. For the final state (halo in equilibrium with
the clump at rest in the centre) we have
\begin{equation} \label{eq:E:f}
  E_{\mathrm{f}} = \frac{\ms}{2}\Phi_{\mathrm{f}}(0)
  - \frac{G}{4}
  \int_0^\infty \frac{M^2_{\mathrm{f}}(r)}{r^2}\,\mathrm{d}r
\end{equation}
-- this is obtained as $E=V/2$ using $M(r)=\ms+M_{\mathrm{f}}(r)$
(assuming a point-mass clump) and ignoring the clump self-energy. If the clump
is extended with cumulative mass profile $\ms(r)$, then
$\Phi_{\mathrm{f}}(0)$ in equation~(\ref{eq:E:f}) has to be replaced by
\begin{equation} \label{eq:Phi0tilde}
  \tilde{\Phi}_{\mathrm{f}0} = -\frac{G}{\ms}
  \int_0^\infty\frac{\ms(r)M_{\mathrm{f}}(r)}{r^2}\,\mathrm{d} r.
\end{equation}
Equating $E_{\mathrm{i}}=E_{\mathrm{f}}$ and re-arranging gives
\begin{equation}\label{eq:energy:balance}
  \ms\left(\varepsilon_{\mathrm{s}} -
  \frac{\Phi_{\mathrm{i}}(r_{\mathrm{i}})+\tilde{\Phi}_{\mathrm{f}0}}{2}
  \right) = \frac{G}{4}\int_0^\infty
  \frac{M^2_{\mathrm{i}}(r)-M^2_{\mathrm{f}}(r)}{r^2} \,\mathrm{d}r > 0.
\end{equation}
This equation relates the clump mass and its initial conditions with the
change in the halo mass profile. Because this latter is a one-dimensional
function, while relation (\ref{eq:energy:balance}) provides only a single
constraint, it can be satisfied by many possible functional forms for the
final mass profile $M_{\mathrm{f}}(r)$. However, using some simple yet
reasonable models for the final mass profile $M_{\mathrm{f}}(r)$ we can obtain
some quantitative estimates for the amount of mass excavated
\begin{equation} \label{eq:Delta:M}
  \Delta M(r) \equiv M_{\mathrm{i}}(r) - M_{\mathrm{f}}(r),
\end{equation}
in particular its maximum and the radius at which it occurs, and their
dependence on clump mass and initial specific energy $\varepsilon_{\mathrm{s}}$.

In order to compare directly to our simulations, we chose the same initial
halo density profile, given in equation~(\ref{eq:rho}) below, as used in the
simulations. For the final halo we assume two different families of models;
the first have mass profile
\begin{equation} \label{eq:A}
  M_{\mathrm{f}}(r) = \left(\frac{r}{x}\right)^{\gamma_{\mathrm{i}}-\gamma_0}
  M_{\mathrm{i}}(r)\quad\text{with $x^\eta=r^\eta+r_0^\eta$},
\end{equation}
where $\gamma_{\mathrm{i}}$ is the central density slope of the initial halo.
These models have a $\rho\propto r^{-\gamma_0}$ density cusp at small
radii. The second family of models also have the central density slope
$\gamma_0$ and scale radius $r_0$ as free parameter and are given by
\begin{equation} \label{eq:B}
  M_{\mathrm{f}}(r) =
  M_{\mathrm{i}}(r)\big(1-\exp(-[r/r_0]^{\gamma_{\mathrm{i}}-\gamma_0})
  \big).
\end{equation}

The top panels of Fig.~\ref{fig:dM} show the resulting relations between clump
mass $\ms$, obtained for the above models from
equation~(\ref{eq:energy:balance}), and the ratio of the maximum excavated
mass over the clump mass, $\Delta M_{\max}/\ms$ for a clump on an orbit with
initial specific energy of $\varepsilon_{\mathrm{s}}\eq-0.12$ (\emph{left},
corresponding to a circular orbit at the halo half-mass radius), or an
initially parabolic orbit ($\varepsilon_{\mathrm{s}}\eq0$, \emph{right})
decaying in a halo with initial density (\ref{eq:rho}), as used in our
$N$-body simulations below. While these two types of models result in a range
of values for the excavated mass, both models suggest that a clump of $1\%$ of
the halo mass can excavate about $1.7$ times its own mass for
$\varepsilon_{\mathrm{s}}\eq-0.12$ and twice its own mass for
$\varepsilon_{\mathrm{s}}\eq0$. The models of equation~(\ref{eq:B}) also
indicate that a less massive clump is relatively more efficient in that it
removes a larger multiple of its own mass.

The radius $r_{\max}$ at which $\Delta M(r)$ becomes maximal (and hence
$\rho_{\mathrm{i}}\eq\rho_{\mathrm{f}}$) is shown in the middle panels of
Fig.~\ref{fig:dM}, again as function of clump mass $\ms$. For a clump of only
1\% of the halo mass, this radius can easily reach the dark-matter scale
radius. We also plot in the bottom panels the radius $r_{50\%}$ inside of
which half the mass has been removed, i.e.\ $M_{\mathrm{f}}(r_{50\%})=
\tfrac{1}{2} M_{\mathrm{i}}(r_{50\%})$, which is readily measured from our
$N$-body simulations below and provides a more direct measure of the size of
any possible density core. The rankings between the models of equation
(\ref{eq:B}) (\emph{solid} curves in Fig.~\ref{fig:dM}) w.r.t.\ $r_{\max}$ and
$r_{50\%}$ are opposite to each other: the model with the smallest $r_{50\%}$
has the largest $r_{\max}$ and $\gamma_0$ and vice versa. While both
$r_{50\%}$ and $r_{\max}$ clearly increase with $\ms$, as one would expect,
this dependence is rather weak, approximately $r\propto\ms^{0.6}$ for the
models of equation~(\ref{eq:B}) near $\ms=0.01M_{\mathrm{halo}}$.

Note that in obtaining these estimates we have assumed that the final halo is
spherical and in perfect equilibrium. In reality, the momentum of the
infalling clump is absorbed by the inner halo, which as a consequence moves
slightly w.r.t.\ its outer parts. This motion is only weakly damped and it
takes some time before its energy is transformed into internal heat. This
implies that the estimates for $\Delta M_{\max}$ from
equation~(\ref{eq:energy:balance}) (as used in Fig.~\ref{fig:dM}) are possibly
somewhat too high, depending on the amount of energy absorbed into such
oscillations. The inner asphericity resulting from the absorbed momentum
also implies that the simple model spherical over-estimates $\Delta M_{\max}$
(because it under-estimates $E_{\mathrm{f}}$ given $M_{\mathrm{f}}$).

\section{Modelling approach}
\label{sec:model}
The above energy argument is only suggestive and cannot predict the final mass
profiles and its dependence on the details of the satellite orbit and the
initial halo equilibrium. To this end numerical ($N$-body) simulations are
required. Recently, \cite{JardelSellwood2009} performed such simulations for a
satellite starting on a circular orbit at the half-mass radius of a spherical
dark-matter halo model with a \cite*{NavarroFrenkWhite1995} density profile and
isotropic velocities. They found that the orbital decay of a satellite with
1\% of the halo mass heats the dark matter and results in a density core of
radius $\sim0.2r_2$, where $r_2$ is the initial scale radius of the halo, the
radius at which $\gamma(r)\equiv-\mathrm{d}\ln\rho/\mathrm{d}\ln r=2$.

A study by \cite{GoerdtEtAl2010} also considered satellites on circular
orbits, but for a variety of halo density profiles with different inner
density slope $\gamma$. These authors were especially interested in the
stalling of the orbital decay \citep{GoerdtEtAl2006,ReadEtAl2006}, which
depends sensitively on $\gamma$.

Assuming a circular orbit for the satellite may simplify some analytical
arguments, but is certainly not very realistic. Similarly, the assumption of
velocity isotropy for the dark matter is not justified and made only for
convenience (so that initial conditions are easily prepared), though
simulations by \cite{ArenaBertin2007} indicate that velocity anisotropy plays
no significant role.

The aim of the present study is to extend the aforementioned simulations to
more realistic initial conditions. In particular, we are interested in
non-circular satellite orbits and cosmologically motivated velocity anisotropy
for the dark matter. Our energy-based argument of \S\ref{sec:analytic} shows
that the total amount of heating induced by the decaying satellite orbit only
depends on the satellite orbit's energy, but not on its eccentricity. However,
where the satellite dumps its energy, and consequently which halo particles
gain energy and angular momentum, will depend on eccentricity. In order to
differentiate these effects, we will present two sets of $N$-body simulations
with satellite orbits of the same energy. The first set essentially extends
the simulations of \citeauthor{JardelSellwood2009} by considering satellite
orbits of varying eccentricity, but same initial radius and energy as used by
those authors. The second set of simulations considers parabolic orbits, whose
initial orbital energy just vanishes. If it were not for dynamical friction,
these orbits would just pass by the dark-matter halo and never
return. However, due to dynamical friction they become bound and decay to the
halo centre if initially aimed sufficiently close. We consider isotropic
velocities for the dark matter as well as velocity anisotropy of various
degree and radial dependence. Furthermore, we investigate the effects from
changing the satellite mass and/or its adopted size.

\subsection{The halo model}
\label{sec:model:halo}
For the density profile of the dark-matter halo, we adopt a truncated
spherical \cite{DehnenMcLaughlin2005} model, which gives an excellent fit to
simulated CDM halos and has density
\begin{equation}\label{eq:rho}
  \rho(r) \propto r^{-7/9}
  \left(r^{4/9}+s^{4/9}\right)^{-6}
  \mathrm{sech}(r/r_{\mathrm{t}}).
\end{equation}
For $r_{\mathrm{t}}\to\infty$, this profile asymptotes to $\rho\propto
r^{-7/9}$ and $r^{-31/9}$ at small and large radii, respectively, with a very
smooth transition. We identify the scale radius with the radius at which
$\gamma(r)=2$ (in the limit $r_{\mathrm{t}}\to\infty$), which for these models
is given by $r_2=(11/13)^{9/4}s\approx0.687s$ and set the truncation radius to
$r_{\mathrm{t}}=10r_2$, which we identify with the virial radius. We consider
various velocity anisotropy profiles; in particular models for which
\begin{equation}\label{eq:beta}
  \beta\equiv 1-\frac{\sigma_{\theta}^2+\sigma_{\phi}^2}{2\sigma_r^2}
\end{equation}
is constant and models for which
\begin{equation}\label{eq:beta:r}
  \beta(r)=\beta_\infty \frac{r^{4/9}}{s^{4/9}+r^{4/9}}.
\end{equation}
These latter models are isotropic in the centre and become increasingly
radially anisotropic (for $\beta_\infty>0$) at large radii, again with a very
smooth transition, and are excellent descriptions of $N$-body CDM haloes
\citep{DehnenMcLaughlin2005}.

To generate initial $N$-body conditions for the halo, we sample positions from
(\ref{eq:rho}) and velocities from self-consistent distribution functions of
the form $L^{-2\beta}f(\varepsilon)$ for constant $\beta$ models with
$f(\varepsilon)$ obtained from an Abel inversion \citep{Cuddeford1991}. For
models with $\beta(r)$ as in equation (\ref{eq:beta:r}), we generate initial
conditions using the made-to-measure $N$-body method of \cite{Dehnen2009}.

For models with constant $\beta$, the resolution in the inner parts is
enhanced by increasing the sampling probability by a factor
$g(\varepsilon)^{-1}$ which is compensated by setting particle masses $\mu_i$
proportional to $g(\varepsilon_i)$. We used
\begin{equation}
  g(\varepsilon) \propto
  \frac{1+q\,r_{\mathrm{circ}}^{4/9}(\varepsilon)}
       {r_{\mathrm{circ}}^{4/9}(\varepsilon)+s^{4/9}}
\end{equation}
with $q{\,=\,}10$ the ratio between maximum and minimum particle mass and
$r_{\mathrm{circ}}(\varepsilon)$ the radius of the circular orbit with
specific energy $\varepsilon$. The gravitational forces were computed using a
softening kernel with density profile given in equation~(\ref{eq:soft:rho})
below and $r_{\mathrm{s}}$ replaced by the softening length $\epsilon=0.005$.
Testing this method for our particular purposes we found that it allows a
reduction of $N$ to half at the same central resolution without any adverse
effects.

We use a unit system where $G=M=r_2=1$, which implies a time unit
of $15.5\,\text{Myr}\, (r_2/500\text{pc})^{3/2}
(M/10^8\mathrm{M}_\odot)^{-1/2}$.

\subsection{Orbital and other parameters of the infalling clump}
\label{sec:model:sat}
Henceforth, we shall use the term `satellite' for the infalling baryonic
clump, which we model as a single massive extended (softened) particle. For
the satellite mass $\ms$ we considered 0.3\%, 1\%, and 3\% of the total halo
mass (only 77\% of which is inside $r_{\mathrm{t}}$ associated with the virial
radius), while the satellite size $\rs$ was taken to be $0.01$, $0.03$, or
$0.1$ times the halo scale radius. This means that the satellite is
effectively modelled to have spherical density profile
\begin{equation} \label{eq:soft:rho}
  \rho_{\mathrm{s}}(r) = \frac{15}{8\pi}
  \frac{\rs^4\,\ms} {(r^2+\rs^2)^{7/2}}.
\end{equation}
We considered a large range of initial satellite orbits, but report here only
on two sets of simulations. The simulations of the first set, summarised in
the top half of table~\ref{tab:results}, all start from $r=4$, the radius
containing 40\% of the total or 54\% of the mass within $r_{\mathrm{t}}$, and
have specific orbital energy $\varepsilon_{\mathrm{s}}=-0.12$ equal to that of
the circular orbit at that radius. The only remaining free parameter of these
orbits is the pericentric radius of the initial orbit---owing to dynamical
friction, the actual trajectory of the satellite may have a slightly smaller
first pericentric radius. These simulations thus extend those reported by
\cite{JardelSellwood2009} to non-zero eccentricity and also anisotropic halo
velocity distributions. The second set of simulations, summarised in the
second half of table~\ref{tab:results}, employs parabolic orbits, i.e.\ with
initial orbital energy $\varepsilon_{\mathrm{s}}=0$, started at
$r=r_{\mathrm{t}}$, corresponding to the halo virial radius. Again, the only
free orbital parameter is the pericentric radius of the initial orbit.

Within either set of simulations the specific energy of the initial orbit is
the same. This choice was motivated by the analytic argument of
section~\ref{sec:analytic}, which suggested that orbital energy is the main
parameter affecting the amount of `damage' done to the halo. Thus keeping this
energy fixed and varying orbital eccentricity or, equivalently, the
pericentric radius, we can study the influence of this secondary orbital
parameter.

\begin{table}
\setlength{\tabcolsep}{0.35em}
\caption{Initial conditions and results for our simulations. Initial
  conditions are specified by the satellite size $\rs$ and mass $\ms$, which
  default to $\rs\eq0.03$ and $\ms\eq0.01$, respectively; the peri-centric
  radius $\rperi$ of the initial satellite orbit; and the halo initial
  velocity anisotropy $\beta_{\mathrm{i}}$, which is either constant or
  $\beta(r)$ given by equation~(\ref{eq:beta:r}) with $\beta_\infty\eq1$.  As
  results we list the time $\tinfall$ for the satellite to fall to the centre
  of the halo (defined in Section~\ref{sec:decay:initial}); the radius
  $r_{\max}$ of maximum halo-mass reduction $\Delta{M}_{\max}$ (compared to
  the control simulation); the radius $r_{50\%}$ where the cumulative mass is
  reduced to 50\% compared to the control simulation; the radius
  $r_{M=m_{\mathrm{s}}}$ within which the final halo mass equals $\ms$; and
  the maximum (over all radii) of $\rho/\sigma^3$ for the final halo.
  \label{tab:results}}
\begin{center}
\begin{tabular}{llr@{}lrrrrrr} 
\hline
\multicolumn{10}{c}{simulations started at $r_{\mathrm{i}}=4$ and run until
$t=1000$} \\[-0.8ex] \hline \\[-3ex]
  \multicolumn{1}{l}{$\rs,\ms$} &
  \multicolumn{1}{c}{$\rperi$} &
  \multicolumn{2}{c}{$\beta_{\mathrm{i}}$} &
  \multicolumn{1}{c}{$\tinfall$} &
  \multicolumn{1}{c}{$r_{\max}$} &
  \multicolumn{1}{c}{$\Delta M_{\max}$} &
  \multicolumn{1}{c}{$r_{50\%}$} &
  \multicolumn{1}{c}{$r_{M=m_{\mathrm{s}}}$} &
  \multicolumn{1}{c}{$\max\Big\{\frac{\rho}{\sigma^3}\Big\}$}\\
\hline
default      &0.00&$-$&0.43 &  44.1& 0.704& 0.0157& 0.330& 0.315& 1.36\\ 
default      &1.03&$-$&0.43 & 111.4& 1.351& 0.0104& 0.119& 0.253& 4.16\\ 
default      &2.44&$-$&0.43 & 169.8& 1.864& 0.0085& 0.100& 0.247& 4.80\\ 
default      &4.01&$-$&0.43 & 198.0& 2.052& 0.0081& 0.109& 0.247& 4.44\\ 
default      &0.00&   &0    &  43.5& 0.646& 0.0139& 0.324& 0.312& 1.14\\ 
default      &1.03&   &0    &  99.9& 0.778& 0.0106& 0.165& 0.266& 2.84\\ 
default      &2.44&   &0    & 163.1& 0.783& 0.0086& 0.100& 0.250& 3.88\\ 
default      &4.01&   &0    & 193.6& 0.779& 0.0084& 0.097& 0.249& 3.57\\ 
default      &0.00&   &0.3  &  43.2& 0.563& 0.0139& 0.350& 0.325& 0.84\\ 
default      &1.03&   &0.3  &  87.1& 0.644& 0.0104& 0.242& 0.287& 1.42\\ 
default      &2.44&   &0.3  & 145.9& 0.641& 0.0088& 0.127& 0.262& 2.46\\ 
default      &4.01&   &0.3  & 178.8& 0.665& 0.0082& 0.128& 0.257& 2.37\\ 
$\rs\,\,{=}0.01$ &4.01&&0   & 181.8& 0.802& 0.0084& 0.048& 0.247& 4.67\\ 
$\rs\,\,{=}0.1 $ &4.01&&0   & 231.1& 0.464& 0.0076& 0.215& 0.272& 2.60\\ 
$\ms{=}0.003$&4.01&   &0    & 491.4& 1.932& 0.0028& 0.042& 0.116&13.38\\ 
$\ms{=}0.03$ &4.01&   &0    &  90.3& 1.278& 0.0210& 0.257& 0.541& 0.88\\ 
\hline
\multicolumn{10}{c}{simulations started at $r_{\mathrm{i}}=10$ and run until
$t=2000$}    \\[-0.8ex]
\hline \\[-3ex]
  \multicolumn{1}{l}{$\rs,\ms$} &
  \multicolumn{1}{c}{$\rperi$} &
  \multicolumn{2}{c}{$\beta_{\mathrm{i}}$} &
  \multicolumn{1}{c}{$\tinfall$} &
  \multicolumn{1}{c}{$r_{\max}$} &
  \multicolumn{1}{c}{$\Delta M_{\max}$} &
  \multicolumn{1}{c}{$r_{50\%}$} &
  \multicolumn{1}{c}{$r_{M=m_{\mathrm{s}}}$} &
  \multicolumn{1}{c}{$\max\Big\{\frac{\rho}{\sigma^3}\Big\}$}\\
\hline
default      &0.0&$-$&0.43   &  225& 0.802& 0.0243& 0.530& 0.384& 0.86\\ 
default      &0.4&$-$&0.43   &  460& 0.891& 0.0208& 0.393& 0.332& 1.26\\ 
default      &0.8&$-$&0.43   &  835& 1.271& 0.0180& 0.207& 0.280& 2.96\\ 
default      &1.3&$-$&0.43   & 1365& 1.922& 0.0163& 0.094& 0.256& 4.32\\ 
default      &0.0&&0         &  215& 0.701& 0.0229& 0.515& 0.378& 0.69\\ 
default      &0.4&&0         &  408& 0.734& 0.0197& 0.447& 0.352& 0.82\\ 
default      &0.8&&0         &  774& 0.906& 0.0167& 0.331& 0.315& 1.20\\ 
default      &1.3&&0         & 1319& 1.288& 0.0138& 0.265& 0.294& 1.62\\ 
default      &0.0&   &0.3    &  202& 0.710& 0.0220& 0.511& 0.380& 0.57\\ 
default      &0.4&   &0.3    &  388& 0.721& 0.0193& 0.445& 0.355& 0.69\\ 
default      &0.8&   &0.3    &  700& 0.750& 0.0178& 0.389& 0.339& 0.88\\ 
default      &1.3&   &0.3    & 1165& 0.853& 0.0160& 0.358& 0.329& 0.84\\ 
default      &0.0&&$\beta(r)$&  183& 0.858& 0.0187& 0.445& 0.372& 0.52\\ 
default      &0.4&&$\beta(r)$&  320& 0.812& 0.0182& 0.435& 0.368& 0.53\\ 
default      &0.8&&$\beta(r)$&  593& 0.807& 0.0157& 0.379& 0.350& 0.66\\ 
default      &1.3&&$\beta(r)$&  947& 0.844& 0.0159& 0.372& 0.348& 0.65\\ 
$\rs\,\,{=}0.01$&0.4&&$\beta(r)$&249&0.834& 0.0174& 0.412& 0.360& 0.58\\ 
$\rs\,\,{=}0.1 $&0.4&&$\beta(r)$&566&0.822& 0.0177& 0.415& 0.360& 0.65\\ 
$\ms{=}0.003$&0.4&&$\beta(r)$& 1902& 0.406& 0.0064& 0.207& 0.180& 2.02\\ 
$\ms{=}0.03$ &0.4&&$\beta(r)$&   95& 1.149& 0.0427& 0.871& 0.762& 0.16\\ 
\hline
\end{tabular}
\end{center}
\end{table}

\subsection{Technicalities}
\label{sec:model:tech}
The $N$-body simulations are performed using the public $N$-body code
\text{gyrfalcON} which uses the $\mathcal{O}(N)$ force solver \textsf{falcON}
\citep{Dehnen2002} with minimum opening parameter $\theta_{\min}=0.5$ and
employs an adaptive time-stepping scheme.

The simulations were run for 1000 or 2000 time units for the first and second
set of simulations, respectively. The energy conservation was typically 1 part
in $10^4$ (more accurate control simulations obtained the same results). Halo
models with constant $\beta$ had 1\,Mio particles selected using the
resolution-enhancement method of Section~\ref{sec:model:halo}, while for halo
models with $\beta(r)$ as in equation~(\ref{eq:beta:r}) 2\,Mio equal-mass
particles were used. In order to ensure a careful modelling of the satellite,
it was integrated with a shorter time step than most halo particles and the
mutual forces with the halo particles were approximated with a much reduced
opening angle. One simulation over 1000 time units (16\,000 block steps or
256\,000 shortest time steps) took about 190 CPU hours (single CPU,
$N=1\,$Mio), including some of the analysis.

\begin{figure*}
  \begin{center}
    \resizebox{84mm}{!}{\includegraphics{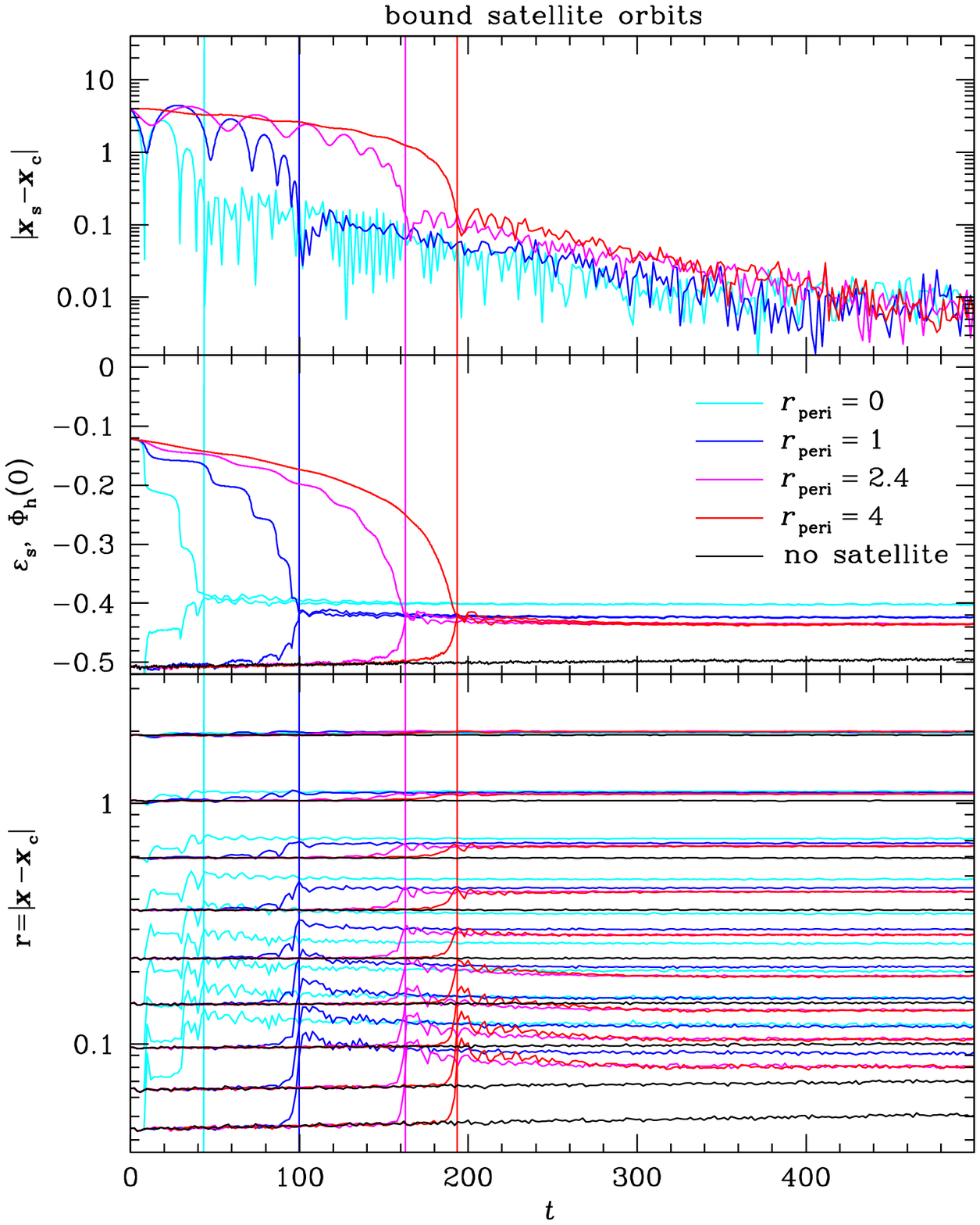}}
    \hfil
    \resizebox{84mm}{!}{\includegraphics{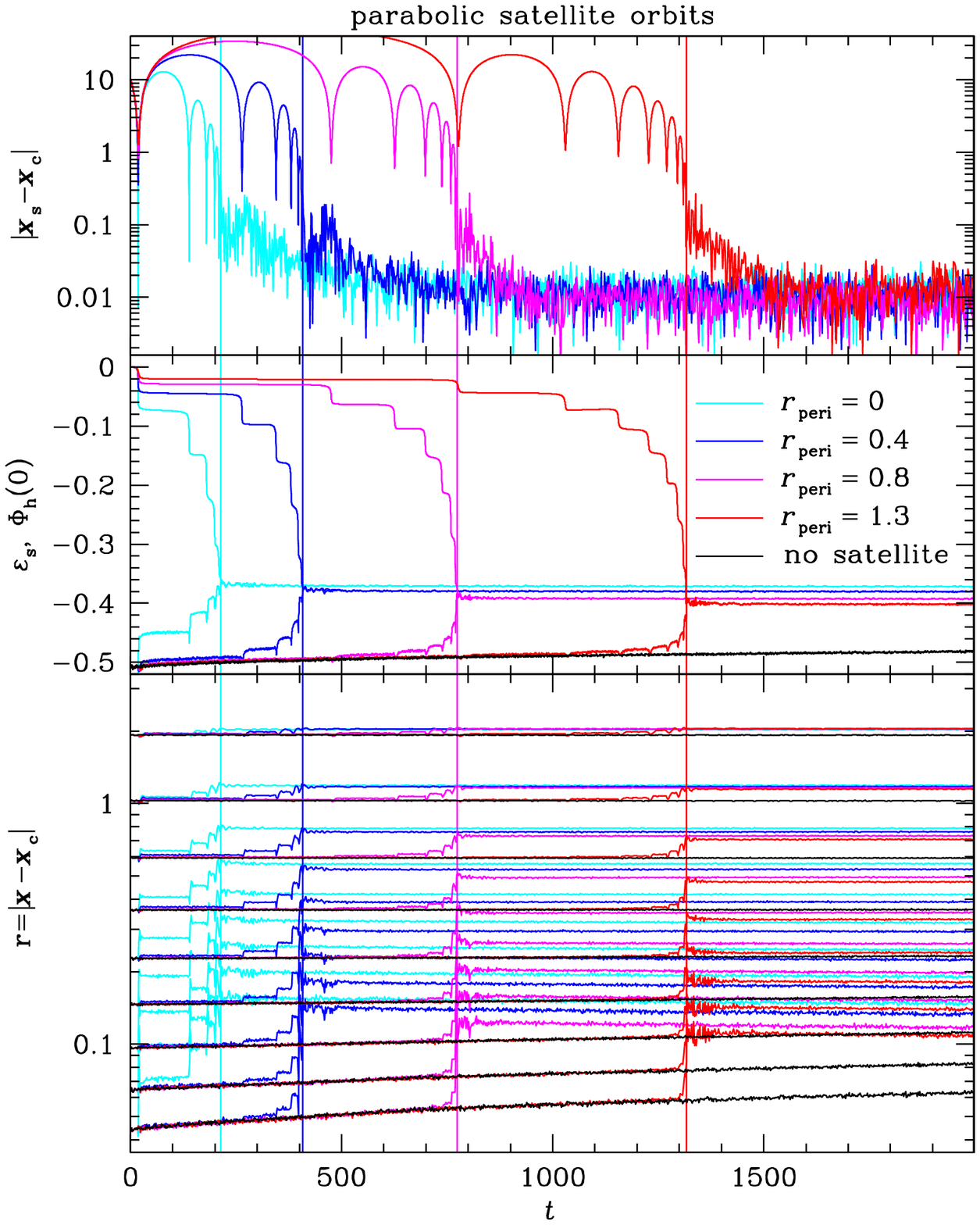}}
  \end{center}
  \caption{
    Time evolution of the satellite orbits and halo Lagrange radii for four
    bound (\emph{left}) and four parabolic satellite orbits (\emph{right})
    decaying in a halo with isotropic velocity distribution -- note the
    different time scales. The \emph{top} and \emph{middle} panels show the
    evolution of the satellite distance to the halo centre and the satellite
    orbital energy $\varepsilon_{\mathrm{s}}$, respectively. Also shown in the
    middle panel is the halo's central potential depth $\Phi_{\mathrm{h}}(0)$
    (lower curves). The thin vertical lines indicate the time at which the
    difference $\varepsilon_{\mathrm{s}}-\Phi_{\mathrm{h}}(0)$ is reduced by a
    factor 50. The \emph{bottom} panel shows the evolution of the halo
    Lagrange radii (w.r.t.\ the halo centre) containing 0.08\%, 0.16\%, etc.~
    up to 20.48\% of the halo mass.  Within each set of orbits the initial
    orbital energy is the same, namely $\varepsilon_{\mathrm{s}}=0$ for the
    parabolic orbits and $\varepsilon_{\mathrm{s}}=-0.12$ for the bound
    orbits, equivalent to that of the circular orbit at the halo half-mass
    radius (which is in fact the \emph{red} orbit in the left panels). The
    purely radial orbits of both sets are plotted in \emph{cyan}, while the
    colour sequence to \emph{red} corresponds to ever less eccentric orbits,
    reaching $e=0$ (circular) for the family of bound orbits, corresponding to
    larger pericentric radii as indicated. \label{fig:decay} }
\end{figure*}
After each time step the position and velocity of the halo centre was
estimated from the position of the most bound particles\footnote{Let $\phi$
  denote a halo particle's specific potential energy due to all other halo
  particles (but not the satellite), then we first find the particle with the
  smallest (most negative) $\phi$ and its $K\eq256$ spatially nearest
  neighbours. Then we obtain the centre position and velocity as weighted
  average from the $K/4$ most bound particles (with largest $|\phi|$) using
  the weights $|\phi_{K/4}-\phi_i|^3$.}. Snapshots are stored at regular
intervals and analysed in terms of their radial profiles using two different
approaches. The first employs simple averages over radial shells, assuming the
centre to coincide with the one found from the most bound particles. The
second estimates for each particle the density as a kernel estimate from its
32 nearest neighbours and then computes the centre, radius, and other
properties from density bins. This latter method is more robust in case the
configuration is non-spherical, either because of flattening or because of a
spatial or velocity offset of the inner w.r.t.\ outer regions.

For all except one halo model, simulations in isolation maintained the
original density profile over 2000 time units, except for the very inner
parts, where artificial two-body relaxation and force softening result in a
slight expansion.  More quantitatively, the Lagrange radii containing
$\lesssim1\times10^{-3}$ of the total mass expand noticably (see bottom panels
of Fig.~\ref{fig:decay}), which is more pronounced in halo models with
radially biased velocity distributions. The exception is the halo model with
initial $\beta(r)$ following equation~(\ref{eq:beta:r}) with
$\beta_\infty=1$. This model turned out to be unstable to the radial-orbit
instability and spontaneously re-arranges into a triaxial configuration within
$\sim\,$200 time units.  In the course of this process the radial mass
distribution is also slightly altered even without infalling satellite. In
order to minimize the effects of these problems when interpreting our results,
we compare each simulation with infalling satellite to a control simulation
without satellite but identical initial halo. Table \ref{tab:results} shows a
numerical summary of our results.

\section{Orbital decay}
\label{sec:decay}
In Fig.~\ref{fig:decay}, we plot the time evolution of the satellite orbital
radius and energy (top and middle panels) and the halo Lagrange radii (bottom
panels) for both sets of satellite orbits in a dark-matter halo with isotropic
velocities. In all these simulations the satellite mass and size are at their
default values of $\ms=0.01$ and $\rs=0.03$.

\subsection{Initial orbital decay} \label{sec:decay:initial}
We like to start our discussion by comparing the circular orbit starting at
$r=4$ (\emph{red} in the left panels), which is similar to that used by
\cite{JardelSellwood2009}, and a plunging, almost radial, orbit with vanishing
orbital energy (\emph{blue} in the right panels of Fig.~\ref{fig:decay}).
While both orbits decay to the centre, their evolution is clearly
different. The circular orbit decays first slowly then faster, whereby
remaining near-circular. The rate of decay, as indicated by the energy loss,
is continuously increasing as the orbit comes closer to the centre. This is
qualitatively consistent with Chandrasekhar's formula~(\ref{eq:dynfr}), since
the halo density is steeply increasing inwards.

The plunging parabolic orbit (\emph{blue} in the right panels) on the other
hand, suffers noticeable energy loss only near peri-centre, undergoing a
stepped rather than steady decay of orbital energy. At the first peri-centric
passage, the satellite loses enough orbital energy to find itself on a bound
but highly eccentric orbit which returns to the inner parts of the halo within
about 4 halo half-mass dynamical times. The energy loss at the second
pericentric passage is larger, resulting in a quick orbital decay thereafter.
The time scale (note the different the time axes in the left and right panels
of Fig.~\ref{fig:decay}) for the decay of this orbit is only about twice as
long as that of the circular orbit starting much closer (a circular orbit
starting at $r_{\mathrm{t}}$ does not decay within 1000 time units). Thus,
even though more energy has to be lost for this orbit, the dynamical friction
at peri-centre is so strong that the decay is still quite fast, even though
the orbit spends about half its time well outside the halo virial radius.

During the first few periods, the peri-centric radius for this plunging orbit
is hardly decaying. This is expected if dynamical friction causes a
near-instantaneous deceleration at peri-centre, which transfers the satellite
to an orbit with the same pericentric radius. The apparent decay of the
peri-centric radius after two periods is not because dynamical friction away
from pericentre becomes significant, as the closest approach to the origin (as
opposed to the halo centre) is in fact increasing, but because the centre of
the halo has moved, as a consequence of the satellite interaction.

For all orbits, we identify as the orbital decay or infall time
$t_{\mathrm{infall}}$ the time at which the difference
$\varepsilon_{\mathrm{s}}-\Phi_{\mathrm{h}}(0)$ first obtains 2\% of its
original value. There is a systematic trend of the infall time to decrease for
higher eccentricities (smaller $\rperi$ within each set of simulations), which
is easily understood by the fact that dynamical friction is stronger for the
higher density at smaller radii.

\begin{figure*}
  \begin{center}
    \resizebox{84mm}{!}{\includegraphics{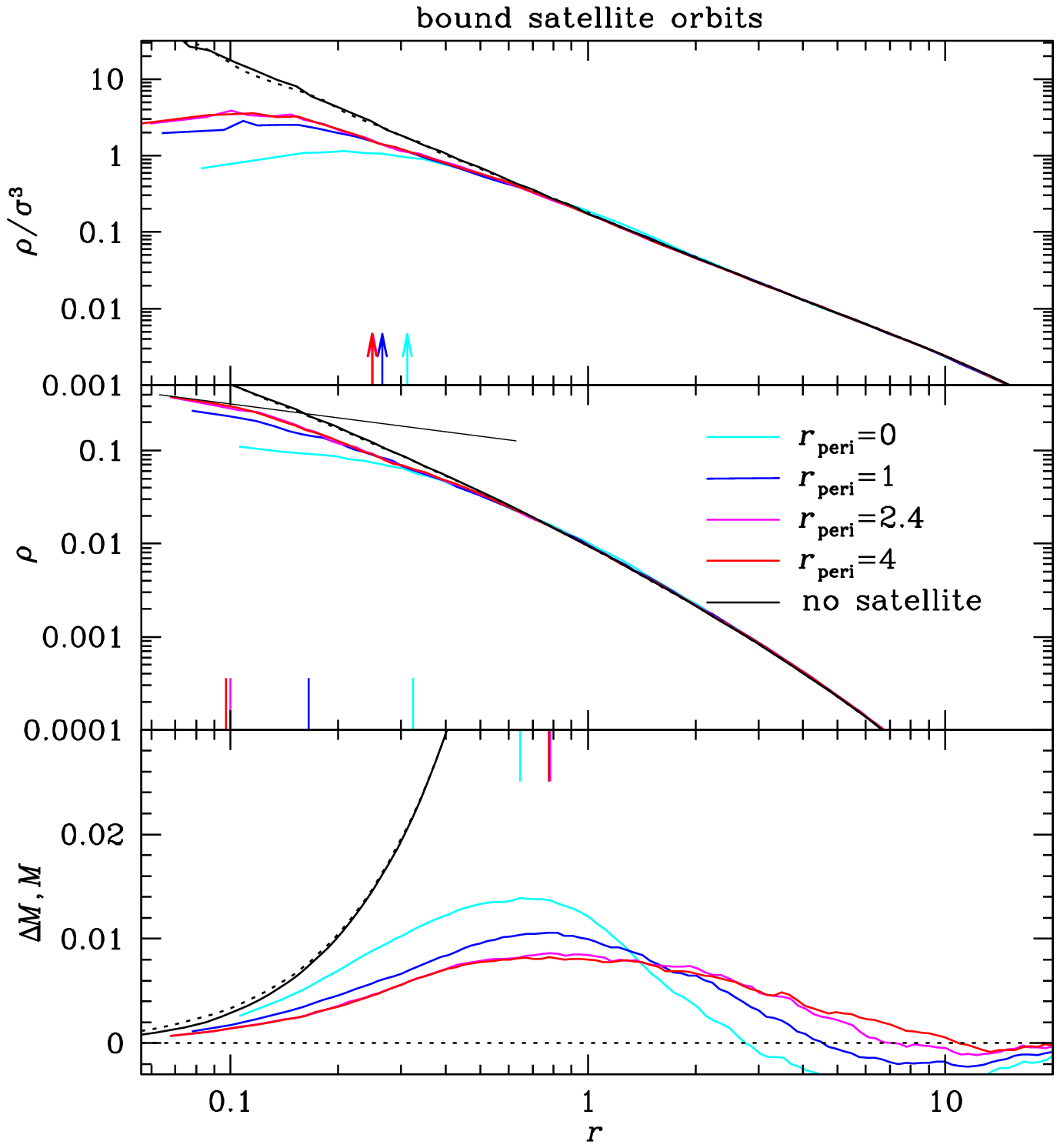}}
    \hfil
    \resizebox{84mm}{!}{\includegraphics{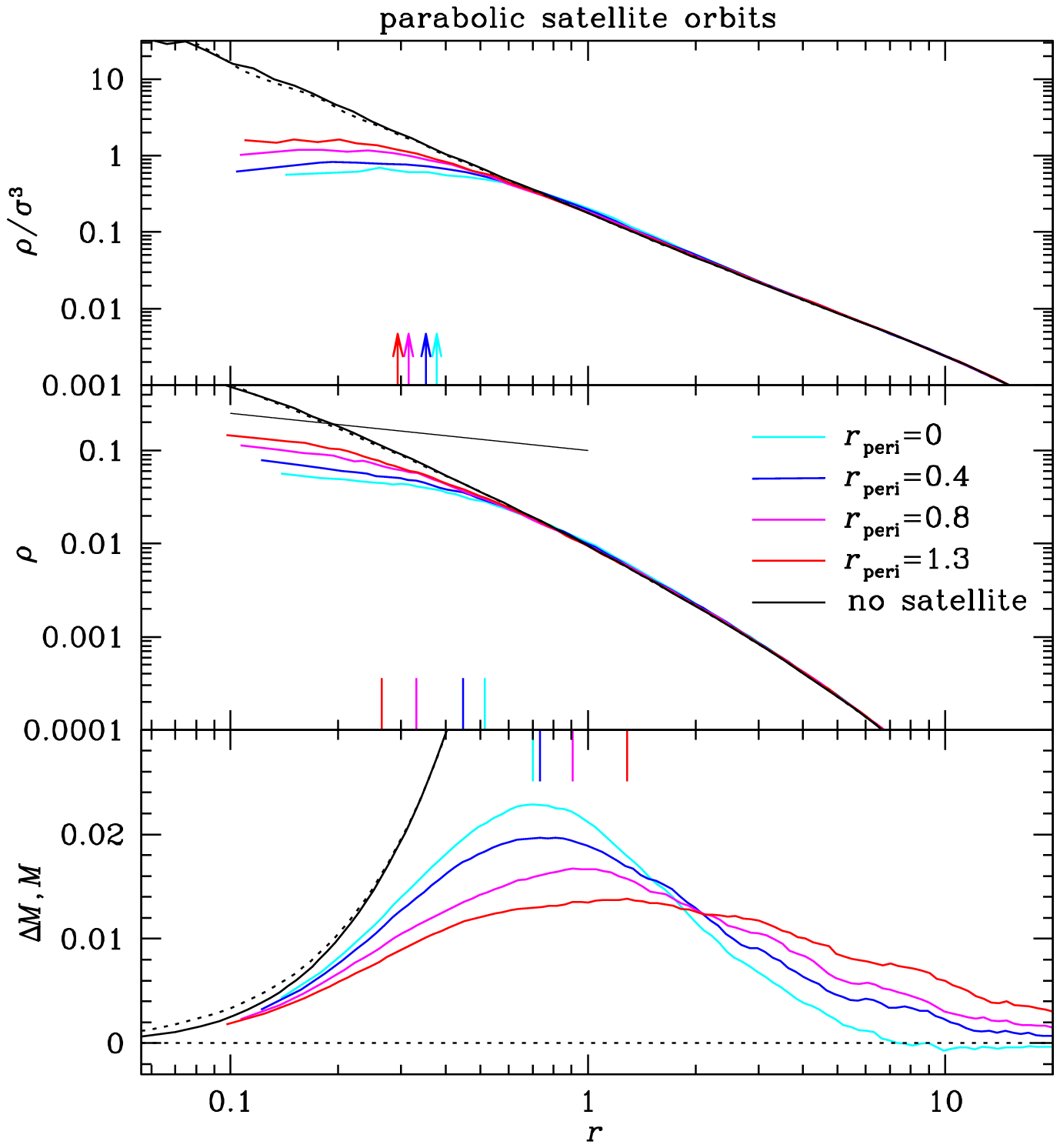}}
  \end{center}
  \caption{\label{fig:profile} Radial profiles of the halo's pseudo
    phase-space density $\rho/\sigma^3$ (\emph{top}), density $\rho$
    (\emph{middle}), and the change in cumulative halo mass $\Delta M$
    (\emph{bottom}) at $t=1000$ and $2000$ for simulations of the decay of
    bound (\emph{left}) and parabolic (\emph{right}) satellite orbits,
    respectively. The simulations and colour coding are the same as in
    Fig.~\ref{fig:decay}. For each model, the vertical lines in the middle and
    bottom panels indicate the locations of $r_{50\%}$ and $r_{\rm max}$,
    respectively, while the arrows in the top panels indicate the radius
    $r_{M=\ms}$ within which the dark mass equals the satellite mass. The
    dotted curves correspond to the situation before satellite infall ($t=0$),
    while the solid black curves represent the control simulation (no
    satellite). In the bottom panels, the black curves give $M(r)$, not
    $\Delta M(r)$. The thin black lines in the middle panels are power-laws
    $r^{-0.5}$(\emph{left}) and $r^{-0.4}$ (\emph{right}).
  }
\end{figure*}

\subsection{Late orbital decay} \label{sec:decay:late}
Interestingly, after the initial decay, when the orbital energy has almost
reached its final value, the satellite and the halo centre are still orbiting
each other at a distance of initially $r\sim0.1$. In other words, the orbital
decay was incomplete and has stalled. Unlike the situation investigated by
previous authors, this stalling occurs at a radius which contains much less
dark mass than the satellite mass $\ms$ and it seems more apt to say the halo
centre orbits the satellite.

This small orbit remains inert only for a short while (when
$|\B{x}_{\mathrm{s}}-\B{x}_{\mathrm{c}}|$ remains constant or even increases
at $\sim0.1$, see also Fig.~\ref{fig:2ndecay}), which is longer for high
eccentricities of the initial orbit, while for the initially circular orbit
($\rperi=4$) there is hardly any such stalling. After this brief pause, the
orbit decays very nearly exponentially with radial e-folding time of $\sim90$
time units for the bound orbits (\emph{left} panels of Fig.~\ref{fig:decay}),
which is discernable for over two e-foldings until the orbital radius reaches
the noise level. In case of initially parabolic satellite orbits
(\emph{right}), the stalled orbits are more eccentric and their decay more
erratic (but qualitatively similar) with a longer decay time (e-folding time
of $\sim160$).

Remarkably, the secondary decay times are very similar between orbits within
either set of simulation, but differ between them. This implies that the
eccentricity of the initial and also the stalled orbit are not affecting the
process responsible for this phenomenon. The differences in secondary decay
times may be caused by differences, induced by the orbital decay, in the
structure of the central halo, as discussed below.

This secondary decay is not associated with any significant orbital energy
loss and has no measurable effect on the final halo density profile (except
perhaps for the innermost 0.1\% of the halo mass), which is the main interest
of our study. However, it is certainly an interesting stellar dynamical
phenomenon deserving further investigation.

\section{Effect on the Halo}
\label{sec:halo:change}
We now discuss the changes induced by the satellite's orbital decay in the
dark-matter halo, concentrating mostly on the spatial distribution, while the
changes to the velocity structure are mostly discussed in the next section.

\subsection{Halo expansion}
The middle panels of Fig.~\ref{fig:decay} also show the evolution of the
central halo potential $\Phi_{\mathrm{h}}(0)$ (lower curves; not including the
contribution of the satellite), which at late time coincides with the
satellite's orbital energy. In all simulations, the final central potential
depth of the halo is considerably shallower than initially, indicating a
significant reduction of the central dark-matter concentration. A comparison
between the various simulations also shows that this reduction is more
pronounced for parabolic than for bound satellite orbits, exactly as our
analytic arguments of Section~\ref{sec:analytic} predicted, as well as for
more eccentric orbits.

In the bottom panels of Fig.~\ref{fig:decay}, we plot the time evolution of
the halo Lagrange radii, while Fig.~\ref{fig:profile} shows for the same
simulations the halo density $\rho$ before and after the satellite infall, as
well as the change (\ref{eq:Delta:M}) in the cumulative halo mass profile and
in the pseudo phase-space density $\rho/\sigma^3$. From the time evolution of
the Lagrange radii, we see that the expansion of the inner halo occurs rather
suddenly at the time when the satellite settles in the core\footnote{The
  initial slow rise of the innermost Lagrange radii is entirely due to
  artificial two-body relaxation and not present in simulations with ten times
  the number of halo particles, see also the last paragraph of
  \S\ref{sec:model:tech}.}, in particular for the initially circular orbit
(\emph{red} in the left panels of Fig.~\ref{fig:decay}). For more eccentric
orbits, the early peri-centre passages result in some minor expansion of the
halo at radii comparable to the peri-centre radius, but hardly affect the
innermost halo.

The mass $\Delta M(r)$ removed from inside radius $r$ (compared to the control
simulation%
\footnote{For the simulations, $\Delta M(r)$ is always measured this way,
  rather than against the initial model (as in equation~\ref{eq:Delta:M}), in
  order to account for halo evolution in absence of any satellite.})  has
different radial profiles for the various simulations. Its amplitude is
generally larger after the decay of a parabolic than a bound orbit, which is
due to the larger orbital energy of the former leading to stronger heating of
the dark-matter particles. In fact, the maximum mass excavated $\Delta
M_{\max}$ is about twice the satellite mass if the latter is decaying on a
plunging parabolic orbit, consistent with our models of
Section~\ref{sec:analytic}, while for a circular orbit decaying from $r=4$,
corresponding to the situation studied by \cite{JardelSellwood2009}, $\Delta
M_{\max}{\,<\,}\ms$. This difference between $\Delta M_{\max}$ obtained for
orbits with $\varepsilon_{\mathrm{s}}=0$ (parabolic)
$\varepsilon_{\mathrm{s}}=-0.12$ (bound) is even more pronounced than for the
analytical models of Fig.~\ref{fig:dM}.

Within each set of orbits $\Delta M(r)$ is most peaked for the purely radial
orbit, while it widens for lower eccentricities.  This is presumably because a
satellite on a highly eccentric loses its energy to a narrow range of
dark-matter particles close to its peri-centre, while less eccentric orbits
lead to a distribution over a wider range.

\subsubsection{Central density reduction}
The density profiles in the middle panels of Fig.~\ref{fig:profile} confirm
earlier results that the infall of a baryonic clump can considerably weaken
the central dark-matter cusp. The final halo mass profiles (not shown) are
noticeably perturbed interior to the radius $r_{\max}$ (indicated by the
vertical lines in the bottom panels of Fig.~\ref{fig:profile}), at which by
definition the final halo density equals that of the control simulation
without satellite. In order to quantify further the properties of the inner
halo density profiles, we also calculate the radius $r_{50\%}$ (indicated by
the vertical lines in the middle panels of Fig.~\ref{fig:profile}) interior to
which the dark mass mass is reduced to 50\% compared to the control
simulation. In general, $r_{50\%}$ is smaller than $r_{\max}$ and marks the
radius at which significant changes to the halo density profile are evident.

There is a clear trend for more eccentric orbits to more strongly reduce the
central density and hence result in a larger radial range over which the
density has been significantly reduced: $r_{50\%}$ is larger for smaller
$\rperi$. The opposite is true for the radius $r_{\max}$. This contrasting
behaviour of $r_{50\%}$ and $r_{\max}$ can be understood in terms of the
different shapes of the excavated mass distribution $\Delta M(r)$, which
becomes broader with increasing $\rperi$, presumably because the energy input
is spread over a larger radial range.

Such an anti-correlation between $r_{50\%}$ and $r_{\max}$ for simulations
with the same initial satellite orbital energy $\varepsilon_{\mathrm{s}}$ was
also present in the analytic models in Fig.~\ref{fig:dM}. For those analytic
models the effect was generated by assuming different central density slopes
$\gamma_0=-\mathrm{d}\ln\rho/\mathrm{d}\ln r|_{r=0}$. However, the central
density slope of the final $N$-body models (middle panel in
Fig.~\ref{fig:profile}) are remarkably similar at around $\gamma_0=0.4-0.5$ at
radii $r\lesssim r_{50\%}$. We should stress, however, that the inner regions
of these halo models are gravitationally dominated by the sunken satellite,
and hence cannot be sensibly compared to the dark-matter haloes of galaxies.

\begin{figure*}
  \begin{center}
    \resizebox{82mm}{!}{\includegraphics{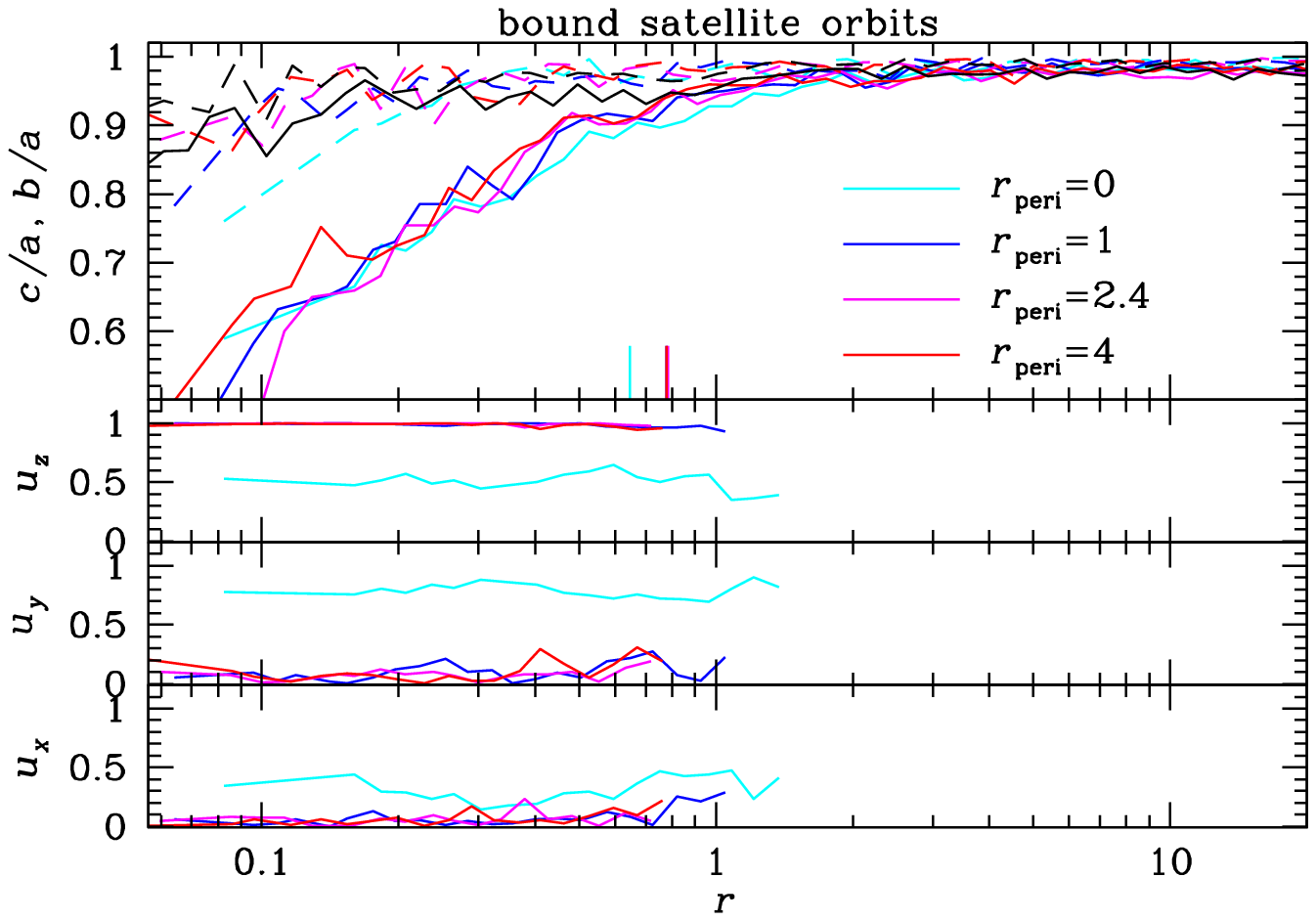}}
    \hfil
    \resizebox{82mm}{!}{\includegraphics{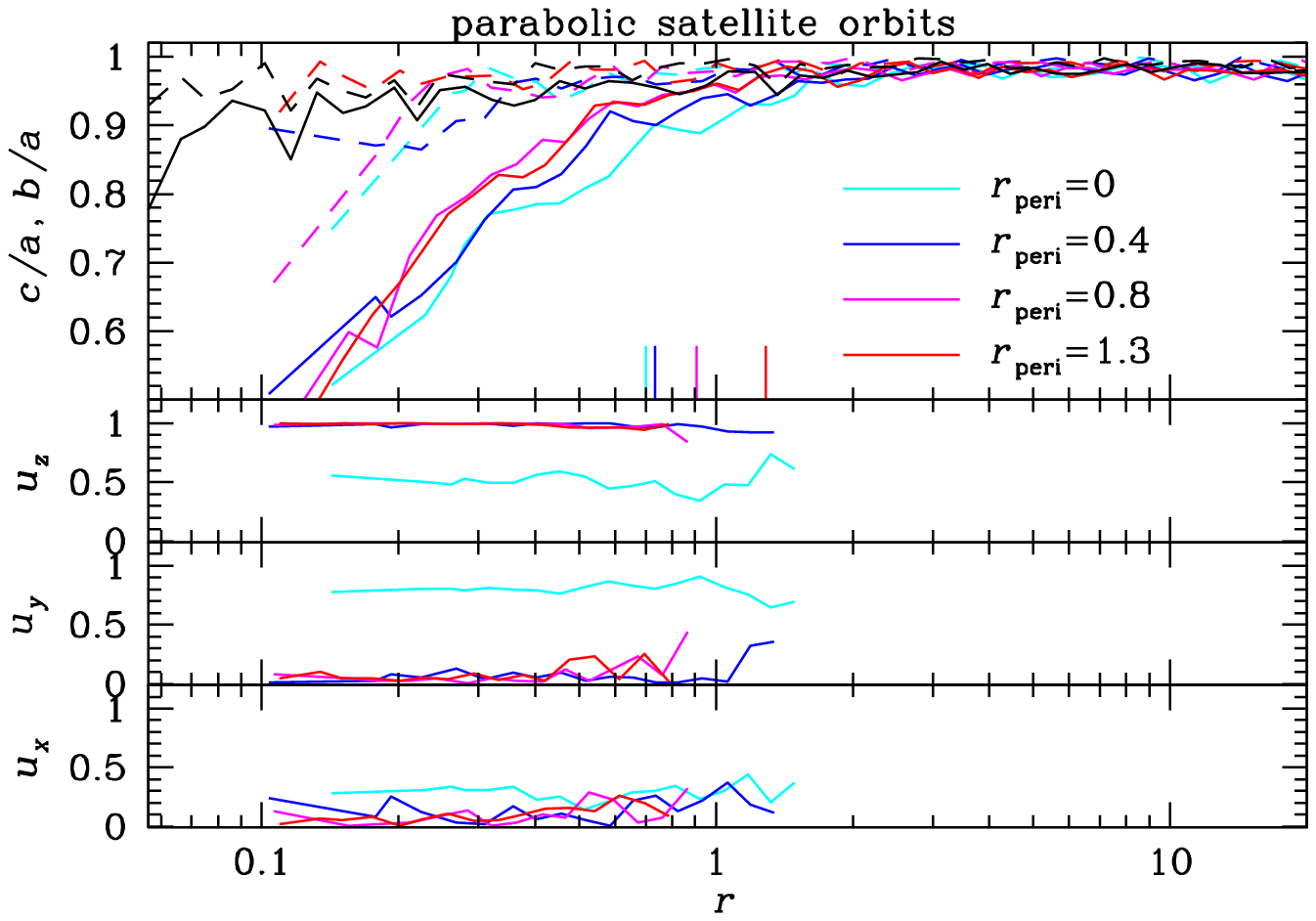}}
  \end{center}
  \caption{\label{fig:shape} Radial profiles of the intermediate-to-major
    ($b/a$, \emph{dashed}) and minor-to-major ($c/a$, \emph{solid}) axis
    ratios, as well as the direction of the minor axis, provided as the $x$,
    $y$, and $z$ components of the unit vector along the minor axis (plotted
    where $c/a<0.95$) for the final halo in the same simulations as in
    Figs.~\ref{fig:decay} and~\ref{fig:profile}. The vertical lines in the top
    panel indicate for each model the location of $r_{\max}$.}
\end{figure*}

\subsubsection{Central phase-space density reduction}
We also show in Fig.~\ref{fig:profile} the radii $r_{M=\ms}$ at which the
enclosed dark mass equals that of the satellite (arrows in the top
panels). Evidently, these radii do not differ much from $r_{50\%}$, implying
that the innermost final dark-matter profiles are not self-gravitating: their
gravity is dominated by the accreted satellite. Since the satellite size
$\rs\eq0.03$ is about ten times smaller than the radius inside of which it
dominates the dynamics, it is effectively acting like a central point mass. A
tracer population orbiting a point mass has a cusp $\rho\propto r^{-3/2}$ if
its phase-space density is constant. However, the simulations have much
shallower central cusps, suggesting that the actual dark-matter phase-space
density has a central depression, decreasing towards the highest binding
energies.

This is also borne out by the plots of the pseudo phase-space density
$\rho/\sigma^3$ in the top panels of Fig.~\ref{fig:profile}: while initially
and in the control simulations (dotted and solid black curves) these follow a
pure power law \citep[this is how these models are actually constructed,
  see][]{DehnenMcLaughlin2005}, the final profiles show a strong central
depression, suggesting a considerable reduction of dark-matter phase-space
density. The corresponding reduction in spatial density is much weaker,
because the available phase space is increased above that from the
self-gravitating cusp by the deeper central potential due to the accreted
satellite.

The central velocity dispersion initially decreases towards $r=0$ as
$\sigma^2{\,\propto\,}r^{\min(\gamma,2-\gamma)}$ for a self-gravitating
$\rho\propto r^{-\gamma}$ cusp. In all simulations, the final halo has
$\sigma(r)$ increasing towards smaller radii (not shown, but evident from the
strong central depression in $\rho/\sigma^3$ compared to that in $\rho$),
inside the radius $r_{M=\ms}$, as is required for any equilibrium system
dominated by a central mass concentration.

\subsection{Halo shape} \label{sec:halo:shape}
In all our simulations, the dark matter halo is initially spherical. The top
panels of Fig.~\ref{fig:shape} show the run of the final halo's principal axis
ratios for the same simulations as in Figs.~\ref{fig:decay} and
\ref{fig:profile}. At $r\lesssim1.5>r_{\max}$ the final dark matter
distribution is near-oblate in all cases, becoming flatter towards the centre
reaching $c/a\sim0.5$ at the smallest measurable radius. The bottom panels
of the same figure show the direction cosines of the minor axis. Note that the
satellite initially orbits in the $xy$ plane, starting at $y=0$ and
$x=r_{\mathrm{i}}$. For all but the purely radial orbits the halo minor axis
is perpendicular to the initial orbital plane of the satellite. This is easily
understood to originate from the transfer of orbital angular momentum from the
satellite to inner halo particles during peri-centric passages, which
presumably also generates the oblate inner halo shape.

For the purely radial satellite orbits (cyan), which start off with zero
orbital angular momentum, the final halo minor axis does not align with the
satellite orbit, but appears to point in some random direction (though it is
self-aligned). This behaviour is counter-intuitive as the initial models for
these simulations are completely symmetric w.r.t.\ the infall axis. However,
such a break of symmetry is the natural behaviour of radial orbits in the
gravitational potentials of a density cusp. This is best explained by
considering the limit of vanishing peri-centre radius for the change in
azimuth $\Delta\phi$ occurring over one radial period from apo-centre to
apo-centre. For a harmonic potential, corresponding to $\gamma\eq0$, the
radial orbit just passes straight through ($\Delta\phi\eq\pi$), while in the
potential generated by a point mass, corresponding to $\gamma\eq3$, the radial
orbit is reflected ($\Delta\phi\eq2\pi$). For mass distributions with
intermediate values of $\gamma$, such as dark-matter haloes, $\Delta\phi$ is
between these two extremes, and the symmetry of the initial orbit is
broken\footnote{Strictly speaking, the radial orbit with $L=0$ is never
  deflected ($\Delta\phi\eq\pi$) as no transverse forces act on it. However,
  in the limit $L{\,\to\,}0$ one gets $\Delta\phi{\,>\,}\pi$ (except for the
  harmonic potential), i.e.\ $\Delta\phi$ is discontinuous at $L\eq0$ or,
  equivalently, $\rperi\eq0$. In our simulations, the halo potential is
  modelled from the softened particle potentials and deviates from the
  power-law form at $r{\,\lesssim\,}\epsilon\eq0.005$, where it becomes
  harmonic and the discontinuity at $\rperi\eq0$ is removed such that
  $\Delta\phi{\,\approx\,}\pi$ for $\rperi\lesssim\epsilon$. However, even the
  simulations with initially purely radial satellite orbits have actual
  $\rperi{\,>\,}\epsilon$ at first passage (see Fig.~\ref{fig:decay}), such
  that the simulated satellite orbit is actually deflected.}.

This deflection of the satellite orbit is compensated by an equal and opposite
momentum to the inner halo, such that the subsequent relative orbit of the two
is no longer radial. The further evolution follows the same pattern as for the
non-radial orbits: the halo flattens perpendicular to the angular momentum
axis of this orbit. Note that for both radial orbit simulations shown in
Fig.~\ref{fig:shape} the minor axis is near-perpendicular to the original
infall direction ($x$-axis). This is explained by the fact that $\Delta\phi$
for a shallow cusp is only slightly larger than the value $\pi$ for the
harmonic potential, implying that the orbit is only weakly deflected from its
original infall trajectory, which is indeed what we see in our simulations.

There is also some hint of triaxiality at small radii, in particular for the
radial bound orbit. This is somewhat surprising as it mostly occurs within the
region where the satellite dominates the enclosed mass and hence the potential
is near-spherical. We suspect that this very central triaxiality is generated
during the secondary orbital decay if the small decaying orbit is eccentric.

\section{Effect of halo velocity anisotropy}
\label{sec:vary:beta}
So far, we have discussed the results from eight simulations, which differed
only in the satellite orbit, but had the same initial halo model with velocity
isotropy. We are now investigating initially anisotropic dark-matter velocity
distributions. Since parabolic orbits are more realistic for infalling clumps,
we restrict our discussion to the four satellite orbits with initially
vanishing orbital energy $\varepsilon_{\mathrm{s}}=0$ and various
eccentricities. Most of our general conclusion are, however, at least
qualitatively also valid for bound orbits.

In addition to the halo model with velocity isotropy, used in
Sections~\ref{sec:decay} and \ref{sec:halo:change}, we consider three
initially anisotropic models. Two of these have constant anisotropy parameter
of $\beta\eq0.3$ and $\beta\eq-0.43$, respectively, which corresponds to the
same level of anisotropy in the sense of $|\ln\sigma_r/\sigma_{\mathrm{t}}|$
(with the tangential velocity dispersion $\sigma_{\mathrm{t}}^2{\,\equiv\,}
[\sigma^2_\phi+\sigma^2_\theta]/2$). The third anisotropic halo model is
motivated by simulations of galaxy halo formation, which generally predict
that the velocity distribution of the dark matter within haloes is outwardly
increasing radially anisotropic \citep{HansenMoore2006}. We use the anisotropy
profile of equation (\ref{eq:beta:r}) with $\beta_\infty=1$. This halo model
is quite different from the others, as it undergoes a radial orbit
instability, and we discuss it in some more detail in
section~\ref{sec:beta:r}.

\begin{figure}
  \centerline{ \resizebox{84mm}{!}{\includegraphics{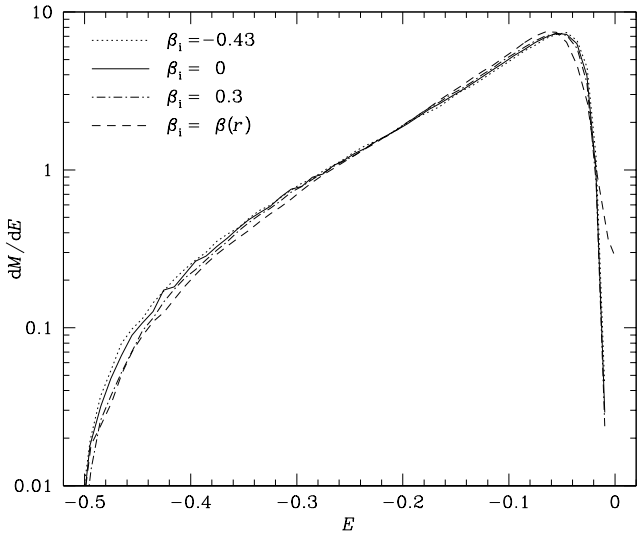}} }
  \caption{
    \label{fig:dMdE}
    Differential energy distributions for the four different halo models
    considered, which only differ in their velocity anisotropy profiles.}
\end{figure}

\subsection{Halo phase-space structure and vulnerability}
\label{sec:beta:vulnerable}
Before we discuss the simulations, let us consider an important difference
between these four models. Haloes with more radially biased velocities have
more mass on eccentric orbits, which spend most of their time near apo-centre,
but contribute significantly to the density near their respective
peri-centres, in particular if the density increases more slowly than $r^{-2}$
(because of geometrical effects) as it does in the inner regions of
dark-matter haloes. Consequently, much of the dark matter in the innermost
region of a halo with radial velocity anisotropy is on orbits which spend most
of their time outside the innermost halo and have lower binding energies than
the local circular orbits. Thus the more radially biased the velocities, the
less mass is at high binding energies, as demonstrated in Fig.~\ref{fig:dMdE},
which shows the differential energy distributions $\mathrm{d}M/\mathrm{d}E$
for our four halo models (computed from $N$-body data; for a cleaner plot of
the same effect but for different models see Fig.~4.5 of
\citeauthor{BinneyTremaine2008}, \citeyear{BinneyTremaine2008}).

The relative lack of highly bound orbits in haloes with radial velocity
anisotropy has immediate consequences for the responsiveness and hence
vulnerability of the central regions to perturbations, such as an infalling
massive satellite \citep[see also][p.~299]{BinneyTremaine2008}. Highly bound
orbits are confined to the central regions, which makes them relatively inert
to external perturbations, in particular if their orbital period is short
compared to the time scale of the perturbation (adiabatic invariance). The
eccentric orbits in haloes with radial velocity anisotropy, on the other hand,
have longer periods and hence are not adiabatically protected in the same
way. Moreover, the infalling satellite can relatively easily perturb
dark-matter particles near the apo-centres of such orbits, increasing their
angular momenta and hence peri-centric radius, and thereby reducing the
central density of the halo.

These arguments also suggest that the difference in vulnerability between
haloes with isotropic and radially anisotropic velocities is more pronounced
for perturbation by a satellite passing at larger peri-centric radius, which
affects eccentric orbits contributing to the centre, but hardly the innermost
orbits. A satellite falling in on a purely radial orbit, on the other hand,
will affect dark-matter orbits at all radii, regardless of halo anisotropy.

\begin{figure}
  \centerline{ \resizebox{84mm}{!}{\includegraphics{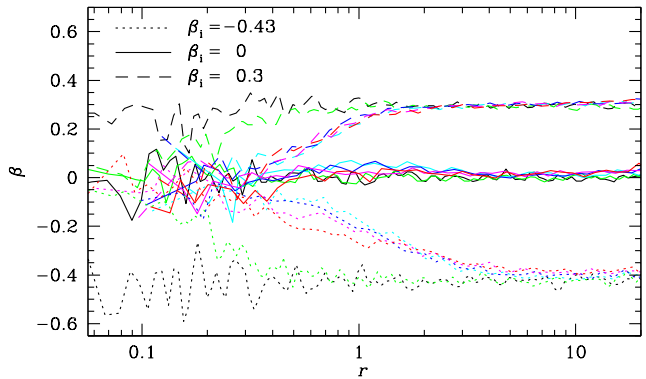}} }
  \caption{
    \label{fig:beta}
    Radial profiles of the final (at $t=2000$) halo velocity anisotropy
    parameter $\beta$ for $4\times3$ simulations: four parabolic satellite
    orbits (as shown in the right panels of Figs.~\ref{fig:decay} to
    \ref{fig:shape} with the same colour coding) decaying in three halo models
    with different constant initial anisotropy (line style as indicated). The
    black and green curves are the radial anisotropy profiles (obtained in the
    same way from $N$-body data) of the initial conditions and the control
    simulations, respectively. The satellite mass and size are $\ms=0.01$ and
    $\rs=0.03$ (as for all simulations presented sofar).}
\end{figure}

\subsection{Change in velocity anisotropy}
In Fig.~\ref{fig:beta}, we plot the radial $\beta$ profiles of the final
halo in simulations of the decay of the four different parabolic satellite
orbits (colour coded as in Figs.~\ref{fig:decay} to \ref{fig:shape}) in each
of the three different halo models with constant initial $\beta$.

In case of initial velocity isotropy the final velocity distribution is
isotropic too, while in all of the initially anisotropic cases, the halo
velocity distribution evolves towards isotropy in the inner regions. This
evolution is partly driven by (artificial) two-body relaxation, as evident
from the control simulations (green in Fig.~\ref{fig:beta}), which results in
velocity isotropy at $r\lesssim0.1$. However, for the simulations with
decaying satellite orbit, this isotropisation occurs to larger radii, which is
essentially independent of the orbital eccentricity of the decaying orbit, and
is only a function of the initial halo anisotropy.

This radius is larger by about a factor 3 for the models with initially
constant tangential velocity anisotropy than for the models with initially
constant radial velocity anisotropy. This is somewhat surprising in view of
the discussion in the previous subsection. One speculation is that a system
with tangential velocity anisotropy is not well-mixed in the sense of
\cite*{TremaineHenonLyndenBell1986} and \cite{Dehnen2005}, such that violent
relaxation, induced by the perturbation, promotes evolution towards isotropy,
while perhaps the opposite is true for radial velocity anisotropy.

It is also interesting to note that the evolution of the anisotropy profile is
complete by the time of the first qualitative change in the orbit
(i.e.\ essentially at $t\eq\tinfall$). Thereafter, the final secondary decay of
the satellite orbit takes place without further modification of the halo
velocity distribution.

\begin{figure}
  \centerline{ \resizebox{84mm}{!}{\includegraphics{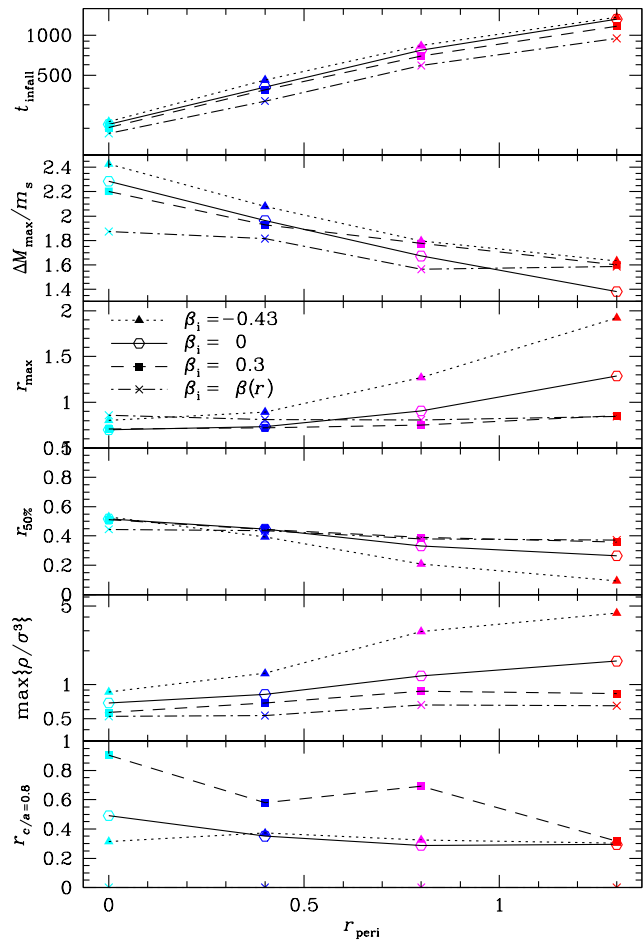}} }
  \caption{
    \label{fig:vary:beta}
    Dependence of orbital decay time $\tinfall$ and several properties of the
    final halo (the bottom panel plots the radius inside of which the
    minor-to-major axis ratio $c/a<0.8$) on initial halo velocity anisotropy
    (line style as in Fig.~\ref{fig:beta}) and initial peri-centre for
    parabolic orbits with satellite of mass $\ms=0.01$ and size
    $\rs=0.03$. Symbol colours match those in Figs.~\ref{fig:decay} to
    \ref{fig:shape}.}
\end{figure}

\subsection{Effect on orbital decay and final halo}
Instead of showing detailed figures, similar to Figs.~\ref{fig:decay} to
\ref{fig:shape}, of the time evolution and radial profiles of density, axis
ratios, etc. for the final halo of all the $4\times4$ simulations, we
summarise the effects of different initial halo anisotropy in
Fig.~\ref{fig:vary:beta}, which shows, for each simulation, the infall time
$\tinfall$ as well as several properties of the final halo (at $t\eq2000$).
Apart from $\Delta M_{\max}/\ms$, $r_{\max}$, and $r_{50\%}$, already used in
previous sections, we also plot the maximum value for the pseudo phase-space
density $\rho/\sigma^3$ and the radius inside of which the minor-to-major axis
ratio $c/a<0.8$.

\begin{figure}
  \centerline{ \resizebox{85mm}{!}{\includegraphics{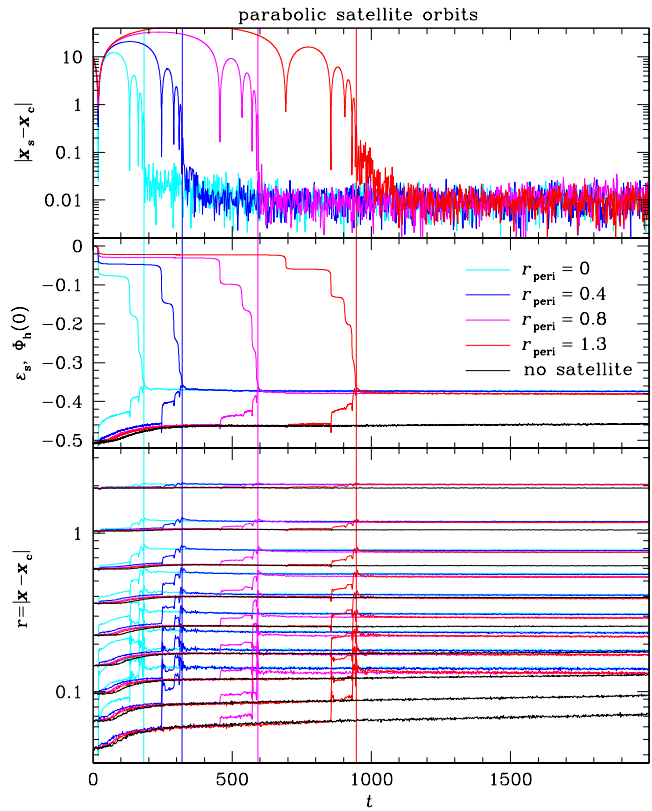}} }
  \caption{
    \label{fig:decay:beta}
    Like the right panels of Fig.~\ref{fig:decay}, but for an initial halo
    with cosmologically motivated velocity anisotropy profile.}
\end{figure}
\begin{figure}
  \centerline{ \resizebox{85mm}{!}{\includegraphics{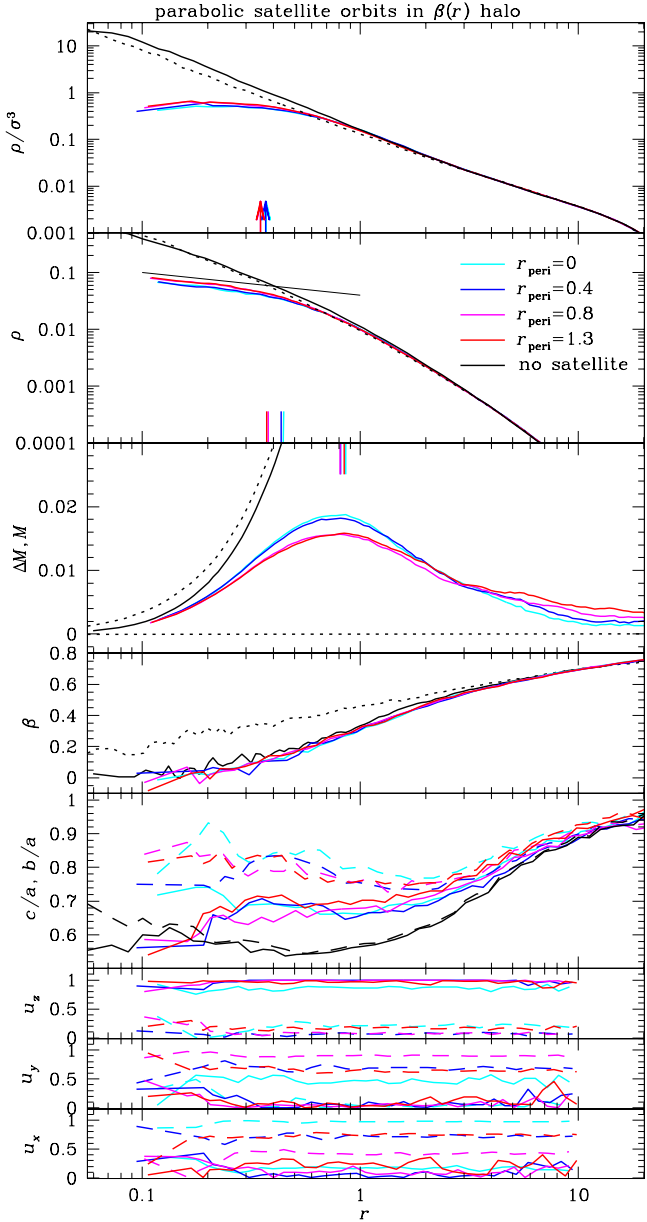}} }
  \caption{
    \label{fig:profile:beta}
    Final halo radial profiles after the infall of a satellite on an initially
    parabolic orbit into a halo with initially radially increasing velocity
    anisotropy $\beta(r)$ as in equation~\ref{eq:beta:r} with
    $\beta_\infty=1$. The plots are equivalent to Figs.~\ref{fig:profile} and
    \ref{fig:shape} (the thin line in the density plot is $r^{-0.4}$), except
    that in the bottom three panels we plot not only the minor-axis direction
    cosines (\emph{solid}), but also the major axis direction cosines
    (\emph{dashed}). For this halo model the control simulation (\emph{black})
    also undergoes some evolution driven by a radial-orbit instability, most
    evident from the changes in anisotropy and shape, see text for details.  }
\end{figure}
There is a systematic trend of shorter infall times for initially more radial
dark-matter velocity anisotropy, exactly as expected from the arguments of
Section~\ref{sec:beta:vulnerable}: the inner parts of haloes with radially
biased velocities are more responsive to and vulnerable by the infalling
satellite. This is also the reason for the behaviour of $r_{50\%}$ and
$\max\{\rho/\sigma^3\}$, which demonstrate that haloes with initially more
radial velocity anisotropy suffer the strongest reduction in their central
density and phase-space density. As argued in
Section~\ref{sec:beta:vulnerable}, the effects of initial halo velocity
anisotropy are most pronounced for satellite orbits with large $\rperi$ and
least for a purely radial satellite orbit.

Again, the radius $r_{\max}$ at which $\Delta M(r)$ peaks anti-correlates with
the radius $r_{50\%}$, indicating that the distribution $\Delta M(r)$ becomes
more peaked (not shown) for more radially biased velocities and that this peak
occurs at smaller radii. Both of these are natural consequences of the
difference in orbital structure as outlined above.

There is also a systematic effect on the halo shape. For each model with
initially constant $\beta$, the bottom panel of Fig.~\ref{fig:vary:beta} plots
the radius inside of which the minor-to-major axis ratio $c/a<0.8$. The halo
with tangential velocity anisotropy becomes flattened in a region comparable
to that of the halo with isotropic velocities. For the halo with $\beta=0.3$,
the flattening is more pronounced, reaching out to about twice as far, and the
shape is near-oblate (not shown) with the same characteristic as for the halo
with initial velocity isotropy, discussed in Section~\ref{sec:halo:shape}.
This is again expected, as the satellite's orbital angular momentum absorbed
by the halo is relatively more important for dark-matter on
low-angular-momentum orbits, as for radial velocity anisotropy.

\section{Halo with cosmologically motivated velocity anisotropy}
\label{sec:beta:r}
The situation with a outwardly increasing radial velocity anisotropy as in
equation~(\ref{eq:beta:r}) is typical for dark-matter haloes emerging from
simulations of large-scale structure formation \citep{HansenMoore2006}. As
already mentioned in section~\ref{sec:model:tech}, our spherical halo model
with such an anisotropy profile undergoes a radial-orbit instability and
quickly settles into a prolate configuration. While this instability somewhat
complicates the interpretation of any results, the corresponding simulations
are presumably most realistic what concerns the halo structure.

\subsection{Orbital Decay}
For the four simulations of initially parabolic satellite orbits in this halo
model, we plot in Fig.~\ref{fig:decay:beta} the orbital decay and time
evolution of the halo Lagrange radii, similar to the right panels of
Fig.~\ref{fig:decay}. First note that the control simulation (black) undergoes
an initial expansion within the first $\sim\,$200 time units. This expansion
is largely driven by the violent relaxation during the re-arrangement to a
prolate configuration (due to the radial-orbit instability), while the
expansion seen in the other control simulations (see Fig.~\ref{fig:decay})
was solely driven by (artificial) two-body relaxation.

The orbital decay is faster than for any other halo model considered (see top
panel of Fig.~\ref{fig:vary:beta}), because of the more efficient transfer of
satellite orbital energy and angular momentum to the halo particles (as
outlined in Section~\ref{sec:beta:vulnerable}) during the very first
peri-centric passages, when the halo density is not yet diminished by the
instability-driven expansion. The associated heating of the innermost halo
results in a slightly faster expansion in the simulations with satellite than
in the control simulation even at $r<\rperi$.

The same is true to a lesser degree for the halo with $\beta=0.3$ (not shown),
while for the halo with isotropic velocities (Fig.~\ref{fig:decay}), the
Lagrange radii at $r<\rperi$ were hardly affected by the early peri-centric
passages. This different response of the innermost halo to the satellite's
first peri-centric passage can be attributed to the predominance of eccentric
dark-matter orbits in the innermost halo, which are perturbed by the passing
satellite at their respective apo-centres, as outlined in
Section~\ref{sec:beta:vulnerable}.

\subsection{Effect on halo structure}
Fig.~\ref{fig:profile:beta} shows the radial profiles of $\rho/\sigma^3$,
$\rho$, $\Delta M$, $\beta$, $c/a$, $b/a$ and the direction cosines of the
minor and major axes (similar to the right panels of Figs.~\ref{fig:profile}
and \ref{fig:shape}) for the four simulations with decaying satellite orbits
(coloured), the initial model (dotted), and control simulation (black).  All
these radial profiles after satellite orbit decay are quite similar, the
strongest difference is 15\% between the amplitudes of $\Delta M(r)$. This
similarity is also expected from our discussion in
Section~\ref{sec:beta:vulnerable}: the halo is so responsive that the exact
satellite orbit does not matter too much.

The pseudo phase-space density, $\rho/\sigma^3$ (top panel), of the control
simulation is even larger than initially, although the density is reduced at
$r\lesssim0.15$. The reason for this counter-intuitive behaviour is that the
isotropisation (evident from the runs of $\beta$) has reduced the velocity
dispersion $\sigma$ in the inner parts. However, after the orbital decay of
the satellite, $\rho/\sigma^3$ is substantially reduced and necessarily also
the true phase-space density. The maximum pseudo phase-space density is a
factor 2-3 smaller than for simulations with initial velocity isotropy, and
the density a factor $\sim2$. This represents the strongest central halo
reduction in all models (for the default values of satellite mass and size),
which makes sense in view of the vulnerability due to the radial velocity
anisotropy.

The mass $\Delta M$ excavated compared to the control simulation is slightly
less for the purely radial orbit (but slightly more for the orbit with
$\rperi=1.3$) than in case of an isotropic halo
(Fig.\,\ref{fig:profile}). However, such a comparison is not quite adequate in
view of the different behaviour of the respective halo models in isolation,
and the picture that stronger radial velocity anisotropy results in larger an
effect on the halo remains valid.

In all simulations is the final halo more isotropic than initially at
$r\lesssim3$, but still retains significant radial anisotropy. In fact, the
$\beta$ profiles are identical to that of the control simulation, presumably
because this profile corresponds to a well-mixed state, attained after the
initial violent relaxation phase, and hardly affected by any further
relaxation due to satellite interaction. Such radially anisotropic velocity
distributions are likely the inevitable property of triaxial and prolate
equilibria, because of the predominance of low-angular-momentum orbits, such
as box orbits \citep{Dehnen2009}.

\subsection{Effect on halo shape}
Obviously, the final halo shapes are completely different from those obtained
in simulations with any other initial halo model. As a result of the
radial-orbit instability, the control simulation obtains a strongly flattened
purely prolate shape with axis ratio $<0.6$ at $r\lesssim2$ (and some random
orientation). This is in remarkable agreement with the fact that dark-matter
haloes in gravity-only simulations of large-scale structure formation tend to
have near-prolate shapes \citep{WarrenEtAl1992}. After the satellite infall,
however, the halo shape becomes less flattened and triaxial in the inner
parts. The changes in halo shape compared to the control simulation extend to
about $r\sim5$, much farther than the changes in halo density or velocity
anisotropy.

As the direction cosines indicate, the orientation of the principle axes of
the triaxial shape is constant with radius, as one expects for an equilibrium
system. The minor axis is always perpendicular to the original satellite
orbit, while the major axis is somewhere in the initial orbital plane (except
for the purely radial orbit for which such a plane cannot be defined).

\begin{figure}
  \centerline{ \resizebox{80mm}{!}{\includegraphics{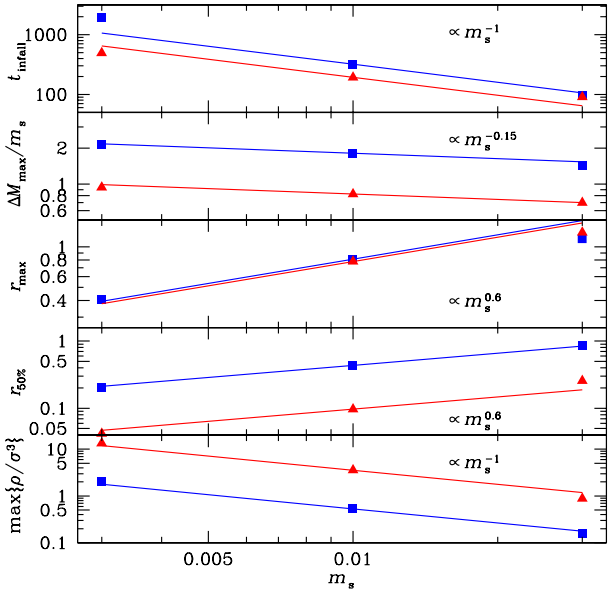}} }
  \caption{
    \label{fig:msat}
    The variation with satellite mass of $\tinfall$, $\Delta M_{\max}$,
    $r_{\max}$, $r_{50\%}$, and the maximum (over all radii) of
    $\rho/\sigma^3$ after the decay of a circular satellite orbit starting at
    $r_{\mathrm{i}}=4$ in a halo with isotropic velocities (\emph{red}) or
    after the decay of a parabolic satellite orbit with $\rperi=0.4$ starting
    at $r_{\mathrm{i}}=10$ in a halo with outwardly increasing radial velocity
    anisotropy (\emph{blue}). The lines are power-laws with exponent as
    indicated.}
\end{figure}

\section{Varying satellite mass and size}
\label{sec:vary:ms:rs}
All simulations presented sofar have the same satellite mass $\ms=0.01$ and
size $\rs=0.03$. We are now investigating effects of larger and smaller values
for these parameters.

\subsection{Varying satellite mass}
Fig.~\ref{fig:msat} shows the effect of varying $\ms$ by a factor of 3 up and
down for a satellite initially on a circular orbit in a halo with isotropic
velocities (\emph{red}) and for a satellite on a parabolic with $\rperi\eq0.4$
in a halo with cosmologically motivated $\beta(r)$. As expected, more massive
satellites reach the centre more rapidly, and cause more damage to the inner
halo. In agreement with Fig.~\ref{fig:dM}, $r_{\max}$, $r_{50\%}$, and $\Delta
M_{\max}$ are largest for the most massive satellite, and decrease
systematically as the satellite mass is decreased. As the figure shows, the
scaling of $r_{\max}$ and $r_{50\%}$ with satellite mass is close to scaling
$r\propto\ms^{0.6}$ (solid lines) also found for our analytic models in
Fig.~\ref{fig:dM}. 

The most interesting result from this study of varying satellite mass is that
low mass satellites are more efficient at displacing mass than high mass
satellites: in agreement with the analytic models the \emph{relative}
excavated mass $\Delta M_{\max}/\ms$ increases towards smaller satellite
masses.

\begin{figure}
  \centerline{ \resizebox{80mm}{!}{\includegraphics{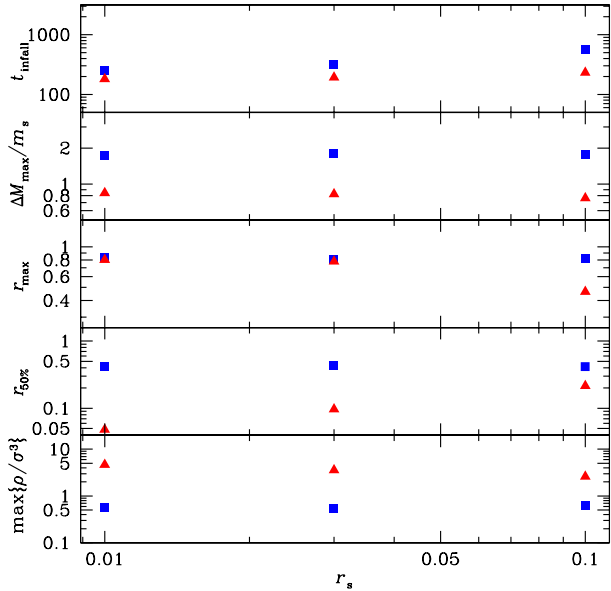}} }
  \caption{
    \label{fig:rsat}
    Like Fig.~\ref{fig:msat}, except that satellite size $\rs$ is varied and
    $\ms=0.01$. kept constant.
  }
\end{figure}
According to Chandrasekhar's dynamical friction formula (\ref{eq:dynfr}), the
drag on the satellite is proportional to its mass $\ms$ and would therefore
naively expect the infall time to scale inversely with satellite mass:
$\tinfall\propto\ms^{-1}$. This is in fact the scaling indicated by the solid
lines in the top panel of Fig.~\ref{fig:msat}. However, the actual infall time
measured for our simulations differ from this expectation. For the simple
situation of an initially circular satellite orbit in a stable halo with
isotropic velocities (red triangles), the infall time scales more like
$\tinfall\propto\ms^{-0.9}$. This difference is most likely caused by the
re-adjustment of the halo during the satellite infall, an effect we ignored
when deducing $\tinfall\propto\ms^{-1}$, which therefore only applies in the
limit $\ms\to0$. The fact that this limit is not applicable even for
$\ms/M_{\mathrm{halo}}=0.01$ may seem surprising, but is not in view of the
fact that only a small fraction of the total halo mass accounts for most of
its binding energy, which in turn enables such a feeble satellite to inflict
significant damage to the halo.

The orbital decay into the halo model with radially increasing $\beta(r)$
(blue squares) is complicated by the fact that the halo simultaneously
undergoes a radial-orbit instability. The slight central density reduction
driven by this instability increases the infall time compared to the original
halo within the first $\sim\,$200 time units. In our simulations, this sets up
a race between satellite infall and instability-driven halo evolution. A more
massive satellite sinks more quickly than this evolution and $\tinfall$ of the
original halo applies, while for a low-mass satellite the longer infall time
after the instability-driven halo evolution applies. Clearly, this race
scenario is an artifact or our simulation setup, and we expect, based on the
arguments given above, that in general $\tinfall$ scales slightly shallower
than $\ms^{-1}$.

\subsection{Varying satellite size}
In Fig.~\ref{fig:rsat} we vary the size $\rs$ of the satellite by a factor 3
up and down, while holding its mass fixed for a circular satellite orbit in an
isotropic halo (red triangles) and a parabolic satellite orbit with
$\rperi\eq0.4$ in a halo with cosmologically motivated $\beta(r)$.
There is a systematic trend that more compact satellites lose energy more
quickly and hence exhibit more rapid orbital decay. This is reasonable as more
compact satellites are more efficient at scattering background particles and
thus lose energy to the halo more rapidly.

The panels of Fig.~\ref{fig:rsat} show that the cumulative effect of the
satellite on the halo is almost independent of its size -- the values of
$\Delta M_{\max}/\ms$, $r_{\rm max}$, and $\max\{\rho/\sigma^3\}$ are
virtually unchanged as the size of the satellite is changed by an order of
magnitude, in particular for the parabolic orbit in the $\beta(r)$ halo. This
is not very surprising in view of the analytical models of
Section~\ref{sec:analytic}: the orbital energy of the satellite is essentially
independent of $\rs$ (as long as $\rs \ll r_2$).

However, for the circular satellite orbit decaying in a $\beta=0$ halo, there
is a systematic trend of larger $r_{50\%}$ with larger $\rs$ and the halo
profile is more significantly flattened by a more extended satellite. This is
also seen in the evolution of the Lagrange radii and is reasonable since the
more extended the satellite, the less energy it loses at large radii and
therefore has more energy to impart to the innermost halo. More extended
satellites on near-circular orbits carry more energy to the inner regions and
hence perturb the halo profile to a greater degree (for highly eccentric
orbits this picture does not apply, as they affect the innermost halo already
at their first peri-centric passage).

The evolution produced by satellites on bound orbits in isotropic haloes is
qualitatively the same as that for parabolic orbits, although the magnitude of
the effects is reduced. In particular, the infall time differs by only 30\%
between the largest and smallest satellites, compared to the factor of three
for the parabolic satellite orbit in the halo with increasing $\beta(r)$.
This again may be caused by the increased vulnerability and responsiveness of
haloes with radial velocity anisotropy, as discussed in
Section~\ref{sec:beta:vulnerable}.

\begin{figure}
  \centerline{ \resizebox{80mm}{!}{\includegraphics{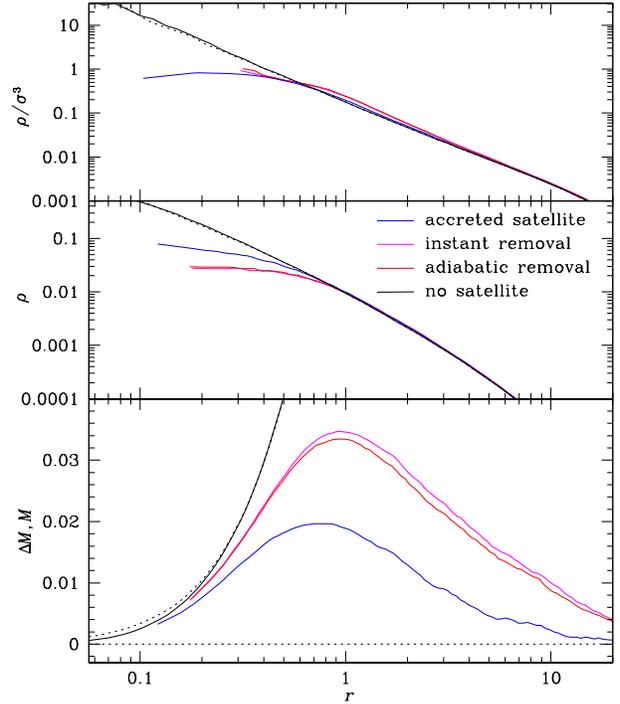}} }
  \caption{
    \label{fig:removal}
    As Fig.~\ref{fig:profile}, but also including the removal of the
    satellite. The \emph{blue} curves refer to the same model as the
    \emph{blue} curves in Fig.~\ref{fig:profile}. The \emph{magenta} curves
    are obtained from this model after instant removal of the satellite; while
    for the \emph{red} model the satelite has been slowly removed.  }
\end{figure}
\begin{figure*}
  \centerline{ \resizebox{120mm}{!}{\includegraphics{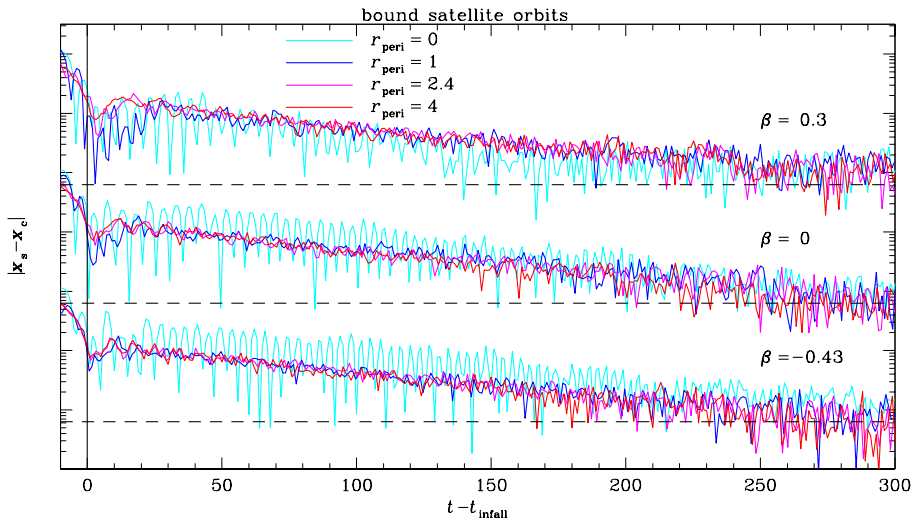}} }
  \caption{
    \label{fig:2ndecay}
    Secondary orbital decay: time evolution of the distance between satellite
    and halo centre for the 300 time units after the infall time (vertical
    line) for the bound satellite orbits in halo models with various velocity
    anisotropy as indicated (the curves for different halo models are offset
    by 2dex from each other). The horizontal dashed lines indicate the noise
    level measured at much later times.  }
\end{figure*}

\section{The Effect of satellite removal}
\label{sec:removal}
In our modelling so far we have ignored the additional halo expansion
following the possible loss of baryons in a feedback-driven galactic wind. At
the final time in the models, the gravitational potential inside $\sim0.1r_2$
is dominated by the satellite. Therefore the central potential is still quite
deep, even though for most of our models the satellite mass is somewhat
smaller than the removed dark mass. This implies that the bound
phase-space volume available for the dark matter at the centre is almost
unchanged.

In order to study the effect of baryon outflow on the dark-matter, we extend
one of our models by removing the satellite. The model concerned is the one
with an initially parabolic satellite orbit with $r_{\mathrm{peri}}=0.4$
falling into a halo with isotropic velocities (\emph{blue} curves in the right
panels of Figs.~\ref{fig:decay} and \ref{fig:profile}). In
Fig.~\ref{fig:removal}, the resulting radial profiles for $\rho/\sigma^3$,
$\rho$, and $\Delta M$ are shown for the situation (i) after satellite
accretion (same as the curves in Fig.~\ref{fig:profile}; \emph{blue}), (ii)
further 1000 time units after instant satellite removal (corresponding to a
fast wind; \emph{magenta}), and (iii) after a slow satellite removal (over 250
times units followed by another 250 time units without satellite; \emph{red}).
Not surprisingly, the satellite removal has a dramatic effect on the central
halo density: the total mass excavated has almost doubled compared to the
situation prior to satellite removal and the density has become clearly cored
with a central density $\sim10$ times smaller than the initial density at the
radius where initially $\ms=M(r)$.

The most intriguing plot, however, is that of the pseudo phase-space density
$\rho/\sigma^3$ (\emph{top} panel of Fig.~\ref{fig:removal}). While the
re-distribution of material to larger radii has shifted the inner profile
somewhat, the overall distribution and the maximum value for $\rho/\sigma^3$
remained unaffected by the satellite removal. For the slow-wind model this is,
of course, expected, since the (true coarse-grained) phase-space density is
conserved. Therefore, the additional reduction of $\rho$ and $\Delta M$ in
this case is entirely owed to the reduction in available phase-space effected
by the removal of the satellite potential. Of course, this is a non-linear
process because the reduction in central dark-matter density itself leads to a
further decrease in the potential depth. This explains why the additional mass
removed by the satellite departure (the difference between the
\emph{red}/\emph{magenta} and \emph{blue} curves) is quite substantial,
exceeding the satellite mass.

Somewhat unexpectedly, the fast-wind model is only slightly more efficient at
removing dark-matter. This strongly suggests that even a fast galactic wind is
nearly adiabatic, i.e.\ has little effect on the (true coarse-grained)
phase-space density (which is unaffected by a slow wind).

\section{Summary}
\label{sec:summary}
We have modelled, using $N$-body simulations, the decay of satellite orbits in
spherical dark-matter halo models with various degrees of velocity anisotropy.

\subsection{Orbital evolution}
The evolution of the satellite orbits has several phases. First there is a
steady decline in the amplitude of the orbit with almost constant peri-centric
radius, driven by dynamical friction near peri-centre. This is followed by a
period of rapid shrinkage of the peri-centre coinciding with significant
expansion of the innermost halo, as evident from the evolution of the Lagrange
radii in Figs.~\ref{fig:decay} and \ref{fig:decay:beta}.

At this stage the evolution of the halo, driven by the transfer of energy and
angular momentum from the decaying satellite orbit, is more or less complete.
The time required for this process scales as $\tinfall\propto\ms^{-0.9}$ with
satellite mass, somewhat shallower than the inverse scaling expected from
Chandrasekhar's dynamical-friction formula. We also find that $\tinfall$ is
significantly shorter for decay in haloes with radial velocity anisotropy,
because the orbital structure of such haloes makes them more susceptible to
perturbations.

The orbital evolution shows some interesting and unexpected behaviour after
$\tinfall$. The most bound region of the halo and the satellite form a sort of
binary, dominated in mass by the satellite, which is modelled as a single
softened particle with size (softening length) $\rs\eq0.03$ (our unit of
length is the halo scale radius). Initially, this secondary orbit has
amplitude $\approx0.1>\rs$, but after a brief period of apparent growth decays
very nearly exponentially with e-folding time of 90 time units.

This is clearly visible in Fig.~\ref{fig:2ndecay}, which plots for all 12
simulations of bound satellite orbits the late time evolution of the distance
$|\B{x}_{\mathrm{s}}-\B{x}_{\mathrm{s}}|$ between halo centre and satellite
after the initial infall of the satellite. In all these cases the secondary
orbital decay occurs. The secondary orbit may be eccentric (as for $\rperi=0$
in Fig.~\ref{fig:2ndecay}) or near-circular, when the decay appears more
regular. In the case of parabolic satellite orbits (not shown), the secondary
decay does not always occur or is much more noisy and less regular.

We are not sure about the cause and dynamics of this phenomenon, and whether
it is a numerical artifact or not. It may be related to the form of the
softened satellite potential, though the secondary orbital amplitude decays
from $\sim0.2>\rs$ to $\sim0.007<\rs$, such that the harmonic inner parts of
the softened potential become relevant only in the late stages.

\subsection{Effect on Halo}
Even though the assumed satellite mass was only 1\% of the total halo mass,
the damage done to the halo is significant: the inner parts of the halo are
substantially reduced in density and phase-space density (as indicated by the
behaviour of $\rho/\sigma^3$), when the satellite has displaced up to twice
its own mass from the innermost halo. The satellite also affects the velocity
structure of the inner halo making it more isotropic. Finally, the halo shape
becomes triaxial in its inner parts.

We find that the efficiency with which the satellite can affect the halo,
e.g.\ the amount of central density reduction, is increased for haloes with
radial velocity anisotropy. This is understandable in terms of the orbital
structure of such haloes, making them more vulnerable (see
Section~\ref{sec:beta:vulnerable}), and relevant, as real dark-matter haloes
are expected, from simulations of large-scale structure formation, to be
radial anisotropic.

One of our most interesting findings is that satellites of smaller mass are
relatively more efficient in damaging the halo: the ratio $\Delta
M_{\max}/\ms$ of the maximum displaced halo mass over the satellite mass
increases towards smaller satellite masses $\ms$ (though the orbital decay
time decreases, of course), as in fact predicted by the analytic energy
arguments presented in Section~\ref{sec:analytic}.

\section{Discussion}
\label{sec:discuss}
In this study we have ignored the details of baryonic physics and instead used
the simple model of a compact baryonic clump falling into a dark-matter
halo. Of all possible scenarios for the effect of baryonic physics on the
structure of dark-matter haloes this is likely the most efficient. This
process proved to be highly effective in altering the structure of the
dark-matter halo in its inner parts, where the galaxy resides. A single clump
of only 1\% of the total halo mass can reduce the density at $\lesssim0.1$
halo scale radii by more than an order of magnitude (including the effect of a
galactic wind as demonstrated in section~\ref{sec:removal}), more than
sufficient to explain the discrepancies between the rotation curves predicted
(using gravity-only simulations) and observed for dark-matter dominated
galaxies \citep[e.g.][]{SimonEtAl2005,deBlokEtAl2008}.

The total amount of baryons is one sixth of all matter
\cite{KomatsuEtAl2010}, i.e.\ 0.2 times the dark mass, 20 times more than
our model clump. Thus, if the baryonic heating of a dark-matter halo is only
$\sim2\%$ efficient on average, the damage done to the halo is equivalent to
that in our models, because any additional addiabatic inflow \emph {and}
outflow of baryons has no lasting effect on the dark matter.

This assumes, of course, that the baryonic sub-structures and clumps are able
to heat the dark-matter particles which contribute to the central cusp. Our
simple model of a compact clump, which sinks to the centre of the dark-matter
halo, is certainly somewhat unrealistic as baryonic structures are susceptible
to disruption by tidal forces before they reach the innermost parts of the
halo. However, owing to their dissipative nature baryons, unlike dark-matter,
can form new sub-structures and clumps, which then continue to heat the
dark-matter via dynamical friction. Moreover, as our simulations have shown,
radial velocity anisotropy, which is typical for dark-matter haloes, make
their innermost parts more vulnerable to perturbations from the outside. This
is because much of the matter in the cuspy region is on eccentric orbits
spending most of their time at much larger radii where they are prone to
dynamic heating (see also Section~\ref{sec:beta:vulnerable}).

\subsection{Dark-matter contraction vs. expansion}
A key process in the re-shaping of dark-matter haloes by non-gravitational
baryonic physics is the transfer of energy via dynamical friction from
baryonic sub-structures to the dark-matter particles\footnote{While this
  process itself is, of course, purely gravitational, it is neglected in
  gravity-only simulations, which ignore the formation of baryonic
  sub-structures and unequivocally predict cuspy dark-matter haloes.}. The
effect of this heating can be understood qualitatively by considering the
Jeans equation of hydrostatic equilibrium
\begin{equation}
  \B{\nabla}(\rho\sigma^2) = - \rho \B{\nabla}\Phi.
\end{equation}
The heating of the dark matter by the non-adiabatic baryon infall raises the
central $\sigma^2$, which, at fixed $\rho$, increases the pressure gradient
(left-hand side). At the same time, the arrival of the baryons increases the
gravitational pull (right-hand side). If these two effects balance exactly,
the dark-matter density $\rho$ remains unaffected. If the heating dominates,
as was the case in our maximally non-adiabatic simulations, then $\rho$ must
flatten to retain equilibrium. Conversely, if the gravitational pull dominates
(especially for negligible heating, i.e.\ `adiabatic contraction'), then
$\rho$ has to steepen (and $\sigma^2$ will increase adiabatically due to the
compression). The exact balance depends on the details and most likely varies
systematically with galaxy type, size, and environment, explaining the
possibility of conflicting results from simulations which attempt to model
baryon physics more directly \citep[e.g.][]{RomanoDiazEtAl2008,
  PedrosaTisseraScannapieco2009}.

An alternative way to look at this problem is to consider the effect on the
dark-matter (coarse-grained) phase-space density, which arguably is
dynamically more relevant than the density, because it is conserved for
adiabatic evolution. However, the process of baryon infall is inevitably
non-adiabatic and reduces the dark-matter phase-space density\footnote{More
  precisely, it reduces the excess-mass function
  \begin{equation}
    D(f) \equiv \textstyle \int_{\bar{f}(\B{x},\B{\upsilon})>f}\,
    \mathrm{d}\B{x}\,\mathrm{d}\B{\upsilon}\,
    \left[\bar{f}(\B{x},\B{\upsilon})-f\right]
  \end{equation}
  with $\bar{f}(\B{x},\B{\upsilon})$ the coarse-grained phase-space density
  \citep{Dehnen2005}.}. As long as the accreted baryons remain at the centre,
this may not necessarily result in a reduction of the dark-matter spatial
density, because the additional gravitational potential of the newly arrived
baryons increases the available bound velocity-space (phase-space at fixed
position).

Of course, a galactic wind changes the situation: the loss of some or all of
the accreted baryons tips the balance towards a reduction of the dark-matter
density, as convincingly demonstrated by our models of satellite removal. In
the Jeans-equation picture the wind removes the additional gravitational pull
from the accreted baryons. In the phase-space interpretation, the wind reduces
the bound velocity space, thus pressing the dark-matter phase-space fluid out
of the centre, like toothpaste out of its tube.

\subsection{Application to dSph and GCs}
For a dwarf spheroidal galaxy our simplistic model of baryon infall may apply
even more directly. Given that their present-day baryonic mass is comparable
to the dark mass that needs to be rearranged in order to convert their haloes
from cusped to cored, it seems plausible that for a reasonable star formation
efficiency one could build the stellar component of the dSphs from a number of
star clusters that fall to the centre by dynamical friction and in so doing
generate a macroscopic core in the halo dark matter distribution of the
dSph. For this to work, the clusters need to remain largely unscathed by
tides before they reach $\sim0.1r_2$. \cite*{PenarrubiaWalkerGilmore2009} have
investigated the tidal disruption of star clusters orbiting in a dSph, in
particular the Sagittarius and Fornax satellites to the Milky Way, and used
$N$-body simulations to confirm the following criterion for the tidal radius
$r_{\mathrm{t}}$ of a globular cluster (GC) at radius $r$ within a halo
\begin{equation}
  \bar{\rho}_{\mathrm{GC}}(r_{\mathrm{t}}) \approx 3\,\bar{\rho}_{\mathrm{h}}(r).
\end{equation}
In particular they looked at the resilience of the five GCs in Fornax and
found that GCs which retain bound masses of greater than approximately 95\% of
their total mass do not undergo tidal disruption by the halo. In our
simulations the clump falls in to approximately $r=0.1r_2$ at which point it
has excavated the mass in the centre of the halo and reduced the central
density. At a corresponding radius in the simulations of
\citeauthor{PenarrubiaWalkerGilmore2009}, four of the GCs are still very
stable against tidal disruption. The fifth one is only disrupted when it
spends a large fraction of a Hubble time at a radius in the range
corresponding to $0.1-0.2r_2$ of our model.

In a similar way we can look at the stability of GCs in Low Surface Brightness
(LSB) galaxies. \cite{KuziodeNarayMcGaughdeBlok2008,KuziodeNarayEtAl2006}
investigated the density profiles of a number of LSB galaxies and attempted to
fit them using NFW and pseudo-isothermal halo models. Their best-fit NFW
models should give an upper limit to the density in the inner halo and
therefore the one most likely able to disrupt an infalling object. Using these
we find a range of values for $\bar{\rho}_{\mathrm{h}}(r)$ at $0.1$ scale
radii ranging from 0.012 to $0.080\,\msun\mathrm{pc}^{-3}$. These are lower
than the corresponding value for Fornax which is
$\sim0.13\,\msun\mathrm{pc}^{-3}$. Based on the work of
\cite{PenarrubiaWalkerGilmore2009} this implies that GCs would also be stable
at $0.1r_2$ in LSB galaxies.

\section*{Acknowledgments}
Research in Theoretical Astrophysics at Leicester is supported by a STFC
rolling grant. MIW acknowledges the Royal Society for support through a
University Research Fellowship. We thank James Binney, Scott Tremaine, and
Justin Read for useful discussions. This research used the ALICE High
Performance Computing Facility at the University of Leicester. Some resources
on ALICE form part of the DiRAC Facility jointly funded by STFC and the Large
Facilities Capital Fund of BIS.


\label{lastpage}
\end{document}